\documentclass[12pt]{article}
\usepackage{epsf,amsmath,amssymb,amsfonts}
\usepackage{graphicx}
\usepackage{epsfig}
\usepackage{verbatim}
\usepackage{cite}
\newcommand{\be}{\begin{equation}}
\newcommand{\ee}{\end{equation}}
\newcommand{\bea}{\begin{eqnarray}}
\newcommand{\eea}{\end{eqnarray}}
\newcommand{\ba}{\begin{eqnarray}}
\newcommand{\ea}{\end{eqnarray}}

\newcommand{\beq}{\begin{equation}}
\newcommand{\eeq}{\end{equation}}
\newcommand{\beqa}{\begin{eqnarray}}
\newcommand{\eeqa}{\end{eqnarray}}
\newcommand{\beqar}{\begin{eqnarray*}}
\newcommand{\eeqar}{\end{eqnarray*}}
\numberwithin{equation}{section}

\begin{document}
\setlength{\unitlength}{1mm}

\thispagestyle{empty} \rightline{\hfill \small hep-th/0511246
 } \rightline{\hfill \small
 ITP-UU-05/54  }
\rightline{\hfill \small  SPIN-05/34 } \rightline{\hfill \small
 ITFA-2005-48 }

\vspace*{2cm}

\begin{center}
{\bf \Large What is the dual of a dipole?}\\

\vspace*{1.4cm}

\vspace*{0.3cm}

{\bf Luis F. Alday,}$^1\,$ {\bf Jan de Boer,}$^2\,$ {\bf Ilies
Messamah}$^2$

\vspace{.8cm}

{\it $^1\,$Institute for Theoretical Physics and Spinoza Institute, Utrecht University \\ 3508 TD Utrecht, The Netherlands}\\[.3em]
{\it $^2\,$Instituut voor Theoretische Fysica,\\ Valckenierstraat
65, 1018XE Amsterdam, The Netherlands.}

\vspace*{0.2cm} {\tt l.f.alday@phys.uu.nl, jdeboer@science.uva.nl,
imessama@science.uva.nl}

\vspace{.8cm}
{\bf Abstract}
\end{center}

We study gravitational solutions that admit a dual CFT description
and carry non zero dipole charge. We focus on the black ring
solution in AdS${}_3\times$S${}^3$ and extract from it the
one-point functions of all CFT operators dual to scalar
excitations of the six-dimensional metric. In the case of small
black rings, characterized by the level $N$, angular momentum $J$
and dipole charge $q_3$, we show how the large $N$ and $J$
dependence of the one-point functions can be reproduced, under
certain assumptions, directly from a suitable ensemble in the dual
CFT. Finally we present a simple toy model that describes the
thermodynamics of the small black ring for arbitrary values of the
dipole charge.

\noindent

\vfill \setcounter{page}{0} \setcounter{footnote}{0}
\newpage

\tableofcontents

\newpage

\setcounter{equation}{0}

\section{Introduction}

One of the surprising features of holography in general and of the
AdS/CFT correspondence in particular is that a local field theory
in $d$ dimensions can look like classical gravity in $d+1$
dimensions. A proper understanding of this relation is crucial in
order to make progress in our understanding of quantum gravity. In
particular, we would like to understand the map between
states/ensembles in the field theory, and smooth/singular
gravitational solutions in the bulk. There are several cases where
we know what this map is. For example, black holes are believed to
be dual to thermal ensembles in the field theory \footnote{There
are several subtleties regarding this statement. For example,
thermal ensembles can be dual to sums over geometries instead of a
single geometry \cite{Maldacena:2001kr}. Also, most pure states in
the thermal ensemble will presumably have a geometrical
description which is similar to that of the black hole in the
regime in which the supergravity description is reliable. From
this point of view the black hole spacetime arises by coarse
graining the underlying microstate geometries. These issues have
little bearing on the results in this paper but will be mentioned
whenever relevant.} as both have a finite periodic Euclidean time.
Other examples include the map between half-BPS states and
geometries in AdS${}_3$ (see
\cite{Lunin:2001jy,mathur2,Mathur:2005zp} and references therein)
and AdS${}_5$
\cite{Lin:2004nb,ma1,ma2,ma3,Balasubramanian:2005mg,ma4}. However,
in general very little is known. We do not even know what
characterizes ensembles that describe general smooth geometries or
geometries with horizons.

Gravitational solutions that are asymptotically AdS have conserved
charges such as mass and angular momentum. These are relatively
easily described in the dual field theory, as for each conserved
charge in the bulk there is a corresponding conserved charge in
the boundary field theory. This is only a very small subset of the
data that characterize the gravitational solution. Since the
latter is asymptotically AdS, it corresponds to a normalizable
deformation of AdS, and this in turn is characterized in full
generality by the one-point functions of all gauge invariant
operators of the dual field theory. It is in principle
straightforward to extract these one-point functions from the
gravitational solution, but the reverse is difficult, as one needs
to integrate the equations of motion of gravity subject to the
boundary conditions specified by the one-point functions.
Similarly, in the boundary theory the relation between one-point
functions and ensembles is known (the one point functions are
simply $\langle {\cal O} \rangle = {\rm tr}(\rho {\cal O})$ with
$\rho$ the density matrix), but to convert one-point functions
into ensembles and vice-versa is practically impossible. Given
one-point functions $\langle {\cal O}_i\rangle$ of operators
${\cal O}_i$, the density matrix that reproduces those expectation
values and has maximal entropy is \be \label{p1} \rho =
\exp(c+\sum a_i {\cal O}_i). \ee Here the coefficients $c$ and
$a_i$ have to be adjusted in such a way that the one-point
functions come out right and such that $\rho$ is properly
normalized. If the set operators ${\cal O}_i$ is a finite subset
of all operators and all other operators have vanishing
expectation value it is not true that (\ref{p1}) should only
involve the operators whose vev is nonzero. This is because it is
by no means obvious that in a density matrix of the form
(\ref{p1}) those other operators will indeed have a vanishing
expectation value, and if this happens it is purely an accident.
Conversely, it may happen that many one-point functions are
nonzero, but all but a finite number of $a_i$ vanish, as for
example in a thermal ensemble. Such ensembles are relatively
simple and one may hope that they have a corresponding simple
gravitational solution, but we have little evidence to support
this beyond what we find in this paper.

In this paper we will try to shed some light on these issues by
considering gravitational solutions that are not just
characterized by conserved charges but also carry a nonzero dipole
charge. This class of solutions typically include giant gravitons
such as in the Myers-Tafjord superstar \cite{Myers:2001aq} in
AdS${}_5$ and black rings in AdS${}_3$ \cite{Emparan:2004wy}.
Dipole charges are not conserved and they will therefore not
correspond to a simple conserved charge in the dual field theory.
On the other hand, they do enter in the generalizations of the
first law for solutions with dipole moments
\cite{Emparan:2004wy,Copsey:2005se}, and as such one would expect
that dipole charges can somehow be incorporated in the dual field
theory. However, the potential that multiplies the dipole charge
in \cite{Emparan:2004wy,Copsey:2005se} involves the difference of
a field at infinity and at the horizon and this field cannot be
put equal to zero at horizon without introducing a singularity at
the rotation axis \cite{Copsey:2005se}. This suggests that dipole
charges should correspond to complicated, probably nonlocal,
operators.

In the case of dipole charges in the 1/2-BPS sector of ${\cal
N}=4$ super Yang-Mills theory this is indeed the case. States in
the 1/2-BPS sector of $U(N)$ ${\cal N}=4$ SYM can be represented,
either via a free fermion or via a free boson representation, in
terms of Young diagrams with $N$ rows \cite{Corley:2001zk,davber}.
In \cite{Balasubramanian:2005mg} it was argued that the number of
giant gravitons appearing in the supergravity solution dual the
state is given by the length of the first row of the Young
diagram. The various giant gravitons carry different dipole
charges (depending on how one defines this notion), so with a
slight abuse of language we will call the number of giant
gravitons the dipole charge of the solution. Strictly speaking we
do the same in the black ring case where the word dipole charge
refers to the number of dipole branes. We hope this will not cause
confusion and keep on calling this the dipole charge of the
system. Thus for the Young diagram the dipole charge is the length
of the first row. This can be reexpressed in terms of the fields
of the free fermion/boson system, but it is easy to see that it
will be rather strange looking non-local operator.

The case of black rings is even less well understood, and these
will be the subject of this paper. Being the first example of
black objects with non-spherical horizon topology, they are also
of interest in their own right. One can take a decoupling limit of
black rings so that they are embedded in AdS${}_3\times$S${}^3$
\cite{Bena:2004tk,{Elvang:2004ds}} and therefore should be
describable in terms of the D1-D5 CFT. This is sometimes called
the UV CFT, not to be confused with the near-horizon IR CFT, the
entropy of the black ring is most easily understood in terms of
the latter. A precise understanding of the black ring in terms of
the UV CFT is more problematic, and a completely convincing
picture has not yet been given. In
\cite{Bena:2004tk} a phenomenological dual description of black
rings is given in terms of a condensate of strings of a fixed
length together with a thermal ensemble of the remaining strings,
employing the familiar short/long string picture of the D1-D5 CFT.
Unfortunately, it is not clear whether this dual picture arises as
a phase in a suitable thermodynamic system, and the size of the
strings in the condensate is fixed by hand by requiring that the
ensemble yields the right entropy and angular momentum.

In the 1/2-BPS limit the situation is somewhat better. 1/2-BPS
black rings, sometimes called small black rings, are characterized
by a central charge $c=6N$ (which is the central charge of the UV
CFT), an angular momentum $J$ and a KK dipole charge $q_3$. There
is a fair amount of evidence that these are dual to ensembles in
the CFT that consist of a Bose-Einstein condensate of $J$ short
strings with length $q_3$ and one unit of angular momentum, plus a
thermal distribution of strings that make up the remaining $N-q_3
J$ units of string
length\cite{{Lunin:2001jy},Bena:2004tk,{Iizuka:2005uv},{Balasubramanian:2005qu}}
. They therefore have a residual entropy of order
$S\sim\sqrt{N-q_3 J}$. The Bose-Einstein condensate only forms for
sufficiently large $J$, typically for $J\sim N$. For $q_3=1$,
which corresponds to the case with no dipole charge, this can be
studied explicitly using a partition function ${\cal Z}={\rm
Tr}(e^{-\beta(H+\mu J)})$ restricted to the 1/2-BPS sector. This
correctly captures the physics of the solution and indeed yields
the picture described above
\cite{don,{Iizuka:2005uv},{Balasubramanian:2005qu},dabholkarrecent}.
One should in principle also be able to further verify this by
comparing the correlation functions of untwisted fields in the
D1-D5 CFT of \cite{Balasubramanian:2005qu} to computations in the
small black ring geometry. The thermodynamic description confirms
that for large $J$ the Bose-Einstein condensate picture is
correct, but for small $J$ it is not a very accurate description
of the system. The thermodynamic description therefore clearly
shows what the generic states carrying fixed angular momentum look
like for different values of the conserved charges.

In the presence of a dipole charge, $q_3>1$, less is known. For
$N=q_3 J$, the small black ring becomes a conical defect as
studied
in\cite{Balasubramanian:2000rt,Maldacena:2000dr,Lunin:2002iz}.
However, a thermodynamic description of such conical defects and
small black rings with $q_3>1$, i.e. one of the form \be
\label{toy} {\cal Z}= {\rm Tr} e^{-\beta(H + \mu J + \nu D)} \ee
with $D$ a ``dipole operator'' is unknown. One of the aims of this
paper was to find such a description. Since this system should
have Bose-Einstein condensation of an excited state (strings of
length larger than one), and not of the ground state, it is clear
that the operator $D$ will have to be rather peculiar to achieve
this.

In this paper we will first study the general black ring in its
decoupling limit in which it is embedded in
AdS${}_3\times$S${}^3$. Using standard AdS/CFT machinery, we will
extract the one-point functions of operators in the dual CFT in
the ensemble dual to the black ring. It turns out that virtually
all operators have non-trivial one-point functions which are
complicated functions of the charges and dipole moments of the
black ring. This is described in section~2. Since it seems rather
hopeless to use these complicated results to extract useful
information about the CFT, we will study 1/2-BPS black rings in
section~3. They can be obtained from the full black ring solution
by taking some charges equal to zero. The one-point functions can
be directly extracted from the results in section~2. These small
black rings have vanishing classical horizon area, but do have a
residual entropy, which might become visible after including
higher order curvature corrections in the supergravity equations
of motion \cite{Dabholkar:2004yr,Iizuka:2005uv}. We notice that it
is possible to choose the charges of the full black ring in such a
way that the solution reproduces both the one-point functions as
well as the entropy of the small black rings. These solutions are
therefore candidates for what the full 1/2-BPS solution in the
presence of higher order corrections could look like. This could
be tested further by studying subleading corrections to the
entropy or one-point functions. It would also be interesting to
understand whether these small extra charges can be understood as
arising from some polarization effect.

In section~4 we turn to the D1-D5 CFT, which we take to be at the
orbifold point. We study one-point functions of operators in the
type of ensembles that are believed to be dual to black rings. In
general, the calculations are way too complicated to perform, as
they involve arbitrarily many twist fields in the orbifold CFT.
Once we restrict to the 1/2-BPS sector the situation improves
somewhat. We argue that in the large $N$ limit correlation
functions simplify and receive only contributions from certain
irreducible pieces. From this we can infer, modulo some
assumptions, what the leading behavior at large $N$ of the
one-point functions is, and we find agreement with the
supergravity analysis. We will also find that one-point functions
in conical defects are generically non-vanishing but are all
suppressed by a single power of $1/N$. This indicates that conical
defect solutions will also receive corrections once higher order
terms in supergravity are included.

These results all indicate that the picture of small black rings
with dipole moments works quite well. Then finally in section~5 we
present a simple toy model of the form (\ref{toy}) which correctly
reproduces the physics of the small black ring. A string of length
$n$ has $H=n$, $J=1$ (if the string carries a unit of angular
momentum), and our proposal is that it carries ``dipole charge''
$D=1/n$. This may seem rather awkward, but it has a couple of
appealing features. First of all, it will not effect the thermal
nature of the partition function at high energies. It therefore
has the flavor of a non-local normalizable deformation, as
expected for a dipole charge. Second, it can be easily generalized
to incorporate more complicated configurations. If, for example,
concentric black rings solutions exist that allow for a decoupling
limit (the solutions of \cite{Gauntlett:2004qy,{Gauntlett:2004wh}}
apparently do not allow such a limit because the charges obey
restrictions incompatible with the decoupling limit) one could
imagine simply including further negative powers of $n$ in the
partition function. Finally, weights of the form $D=1/n$ appear in
many places in integrable systems. Amazingly, exactly the same
operator also appears \cite{Alday:2003zb} by considering the
pp-wave limit of the first non-local conserved charge of string
theory in an AdS background found in \cite{Bena:2003wd}. Clearly,
it would be interesting to explore this connection in more detail.
Some further discussion can be found in the conclusions.

\section{One point functions in SUGRA}

\subsection{The solution}

In \cite{Elvang:2004ds,Bena:2004tk} supergravity solutions
corresponding to supersymmetric black rings with three charges and
three dipoles were obtained. In ten dimensions the solution can be
realized as D1-D5-P black supertubes, carrying the usual charges
of the D1-D5-P system. In addition the solution carries dipole
charges of $D1$ and $D5$ branes as well as KK-monopoles. The
distribution of the various charges is shown in the following
table
\begin{equation}
\begin{array}{llccccccl}
Q_1  &\mbox{D5:} \,\, & z  & z^1  & z^2  & z^3  & z^4  & \_   & \, \\
Q_2  &\mbox{D1:} \,\, & z  & \_ & \_ & \_ & \_ & \_   & \, \\
Q_3  &\mbox{P:}  \,\, & z  & \_ & \_ & \_ & \_ & \_   & \, \\
q_1  &\mbox{d1:} \,\, & \_ & \_ & \_ & \_ & \_ & \psi & \, \\
q_2  &\mbox{d5:} \,\, & \_ & z^1  & z^2  & z^3  & z^4  & \psi & \, \\
q_3  &\mbox{kkm:}\,\, &(z) & z^1  & z^2  & z^3  & z^4  & \psi &
\,.
\end{array}
\end{equation}
The D5-branes wrap a four torus parametrized by $z^1,...,z^4$,
$\psi$ parametrizes a contractible circle (the direction of the
ring) and the solution carries momentum in the $z$-direction,
which is the coordinate that describes the $U(1)$ fiber of the
KK-monopoles. The string frame metric of the black supertube is
given by
\begin{eqnarray}
\label{metricsol}
ds^2
&=& -(X^3)^{1/2} ds_5^2+(X^3)^{-3/2}(dz+A^3)^2+X^1(X^3)^{1/2}dz_4^2 \nonumber\\
&=& -\frac{1}{H_3 \sqrt{H_1
H_2}}(dt+\omega)^2+\frac{H_3}{\sqrt{H_1 H_2} }(dz+A^3)^2+\sqrt{H_1
H_2} dx_4^2+\sqrt{\frac{H_1}{H_2}}dz_4^2 \end{eqnarray} with $X^i=
H_i^{-1} (H_1 H_2 H_3)^{1/3}$, and the harmonic functions $H_i$
are given by \be H_1=1+\frac{Q_1}{\Sigma}-\frac{q_2 q_3 R^2 \cos{2
\theta}}{\Sigma^2},\hspace{0.2in}H_2=1+\frac{Q_2}{\Sigma}-\frac{q_1
q_3 R^2 \cos{2
\theta}}{\Sigma^2},\hspace{0.2in}H_3=1+\frac{Q_3}{\Sigma}-\frac{q_1
q_2 R^2 \cos{2 \theta}}{\Sigma^2} \ee where $\Sigma=r^2 +R^2
\cos^2 \theta$. If we further define \be Y=q_1 Q_1 + q_2 Q_2 + q_3
Q_3 - q_1 q_2 q_3\left(1+\frac{2R^2\cos 2\theta}{\Sigma}\right)
\ee then the one form $\omega = \omega_\phi d\phi+ \omega_\psi
d\psi$ and gauge potential $A^i$ are given by
\begin{eqnarray}
\omega_\phi & = &  - \frac{r^2 \cos^2\theta}{2\Sigma^2} Y \\
  \omega_\psi & = &
    -(q_1+q_2+q_3)\frac{R^2\sin^2\theta}{\Sigma}
    - \frac{(r^2+R^2) \sin^2{\theta}}{2\Sigma^2} Y\\
A^i & = & H_i^{-1}(dt +\omega) +
 \frac{q_i R^2}{\Sigma} (\sin^2\theta d\psi -\cos^2\theta d\phi )
\end{eqnarray}
Finally, the base space $dx_4^2$ is flat space written in a
peculiar coordinate system
\begin{equation}
dx_4^2=\Sigma\left(\frac{dr^2}{r^2+R^2}+d\theta^2\right)
  + (r^2+R^2) \sin^2{\theta}d\psi^2+r^2\cos^2{\theta}d\phi^2
\end{equation}
and $dz_4^2$ is the metric of the four-torus and will play no role
 in the following discussion.  The coordinates take value in the
 range $0 \le r < \infty$, $0 \le \theta \le \pi/2$, $\phi$ and
 $\psi$ have period $2\pi$ and $z$ has period $2\pi R_z$.
The solution also contains a non-vanishing dilaton and RR 3-form
field strength
 \begin{equation}
 e^{2 \Phi}=\frac{H_2}{H_1},\hspace{0.3in}F^{(3)}=(X^1)^{-2} *_5 F^1
 + F^2 \wedge(dz+A^3)
 \end{equation}
where $F^i=dA^i$, and the Hodge dual $*_5$ is with respect to the
metric $ds^2_5$ appearing in (\ref{metricsol}). The solution
depends on 7 parameters, the radius $R$, the charges $Q_i$ and the
dipole charges $q_i$; note that the $Q_i$ are conserved charges at
infinity but the dipole charges $q_i$ are not. These charges are
quantized and related to the integer number of D-branes, momentum
units and dipole branes through
\begin{eqnarray}
\quad N_\mathrm{D5}=\frac{1}{g_s \ell_s^2}\, Q_1\,,\qquad
N_\mathrm{D1}=\frac{1}{g_s
\ell_s^2}\left(\frac{\ell}{\ell_s}\right)^4\, Q_2\,, \qquad
N_\mathrm{P}=\frac{1}{g_s^2
\ell_s^2}\left(\frac{R_z}{\ell_s}\right)^2
\left(\frac{\ell}{\ell_s}\right)^4\, Q_3\,, \nonumber \\
n_\mathrm{D1}= \frac{1}{g_s \ell_s}
\left(\frac{R_z}{\ell_s}\right)\left(\frac{\ell}{\ell_s}\right)^4\,
q_1\,,\qquad n_\mathrm{D5}=
\frac{1}{g_s\ell_s}\left(\frac{R_z}{\ell_s}\right)\, q_2\,,
\label{integercharge}
\end{eqnarray}
with $g_s$ the string coupling constant, $\ell_s$ the string
length and $\ell$ the radius of the torus. Furthermore the
KK-dipole is also quantized in units of $R_z$,
\begin{equation}
q_3=n_{KK} R_z
\end{equation}
for some integer $n_{KK}$. It is useful to define
\begin{equation}
{\cal Q}_1=Q_1-q_2 q_3, \hspace{0.2in}{\cal Q}_2=Q_2-q_1 q_3,
\hspace{0.2in}{\cal Q}_3=Q_3-q_1 q_2
\end{equation}
so that $H_i \ge 0$ implies ${\cal Q}_i \ge 0$. It was shown in
\cite{Elvang:2004ds} that for the solution to be free of closed
causal curves there is also an upper bound for $R^2$ which must be
satisfied
\begin{eqnarray}
\label{noccc} 2 \sum_{i<j} {\cal Q}_i q_i {\cal Q}_j q_j -\sum_i
{\cal Q}_i^2 q_i^2 \ge 4 R^2 q_1 q_2 q_3 \sum_i q_i
\end{eqnarray}
The solution possesses angular momenta in the $\psi$ and $\phi$
directions
\begin{eqnarray}
J_\phi=\frac{1}{2}\frac{R_z \ell^4}{\ell_s^8 g_s^2}(\sum_i q_i Q_i -q_1 q_2 q_3)\\
J_\psi=R^2 \frac{R_z \ell^4}{\ell_s^8 g_s^2}(q_1+q_2 +q_3)+J_\phi
\end{eqnarray}
which need to be integer (or half-integer). Black rings have a
horizon and an associated Bekenstein-Hawking entropy, proportional
to the horizon area. It is most naturally expressed in terms of
the quantized charges; to do so we will denote the integer charges
to which $Q_i$ and $q_i$ are proportional by $N_i$ and $n_i$
respectively. Futhermore, we define $\bar{N}_1=N_1-n_2 n_3$
etcetera, in analogy with our definition of ${\cal Q}_i$. The
entropy is then equal to
\begin{equation}
\label{fullarea} S_{H}=2 \pi \sqrt{n_1 n_2 \left( {\bar N}_1{\bar
N }_2-n_3 J \right)-\frac{1}{4}({\bar N}_1 n_1+{\bar N}_2
n_2-{\bar N}_3 n_3)^2}
\end{equation}
where we have defined $J \equiv J_\psi-J_\phi$.

\bigskip

In the decoupling limit the geometry near the core decouples from
the asymptotically flat region. In order to achieve this
$\alpha'=l_s^2$ must be sent to zero keeping the energies of
excitations near the core as well as the number of branes fixed,
so in this limit
\begin{eqnarray}
r \sim \alpha',\hspace{0.2in} Q_{1,2} \sim \alpha' ,\hspace{0.2in}
Q_3 \sim \alpha'^{\,2}, \hspace{0.2 in} R_z \sim 1\\
R \sim \alpha',\hspace{0.2in} q_{1,2} \sim \alpha', \hspace{0.2in}
q_3 \sim 1, \hspace{0.2in} \ell^2\sim\alpha'
\end{eqnarray}
Note that the region of parameters $Q_1 = Q_2 =Q_3$ and
$q_1=q_2=q_3$, corresponding to the first supersymmetric black
ring found in \cite{Elvang:2004rt}, is not captured by the
decoupling limit.  \footnote{In particular, due to the same
reason, we expect that the multi-centered black rings found in
\cite{Gauntlett:2004qy,Gauntlett:2004wh} do not allow for a
decoupling limit.}

The decoupled solution has the same form (\ref{metricsol}) with
\begin{eqnarray}
H_{1,2}&=&\frac{Q_{1,2}}{\Sigma}-\frac{q_{2,1}q_3R^2\cos
2\theta}{\Sigma^2},\nonumber \\
H_3&=&1+\frac{Q_3}{\Sigma}-\frac{q_1q_2R^2\cos
2\theta}{\Sigma^2},\\
  \omega_\psi &=&
    -q_3\frac{R^2\sin^2\theta}{\Sigma}
    - \frac{(r^2+R^2) \sin^2{\theta}}{\Sigma^2}
    \left(J_{\phi}- q^3\frac{R^2\cos 2\theta}{\Sigma}\right)
  \, .\nonumber
\end{eqnarray}
while $\omega_\phi$ etc have the same form as in the full
solution. To explore the asymptotic region of this metric we first
make a change of variables $z \rightarrow z-t$, after which for
large values of $r$ the metric and three form become (to be
precise what we do is to re-scale $r \rightarrow r/\epsilon, z
\rightarrow \epsilon z, t \rightarrow \epsilon t$ and then take
$\epsilon \rightarrow 0$ )
\begin{eqnarray}
\label{zeroorder} ds^2 & = & \frac{r^2}{\sqrt{Q_1 Q_2
}}(dz^2-dt^2)+\frac{\sqrt{Q_1 Q_2}}{r^2} dr^2 \nonumber +
\sqrt{Q_1 Q_2} (d\theta^2 +\cos^2{\theta} d\phi^2+\sin^2{\theta} d\psi^2)\\
F &  = &  \frac{2r}{\sqrt{Q_1 Q_2}} dz \wedge dt \wedge
dr-\sqrt{Q_1 Q_2} \sin{2 \theta} d\theta \wedge d\psi \wedge
d\phi.
\end{eqnarray}
We see that the asymptotic geometry is identical to that of $M=0$
BTZ black hole times $S^3$, which at large $r$ is asymptotically
$AdS_3 \times S^3$, with radius $(Q_1 Q_2)^{1/4}$. So the black
ring\footnote{Actually, once we uplift to six dimensions the
solution is more properly thought of as a black supertube instead
of a black ring, but we will continue to call the solution a black
ring anyway, hoping that this will not cause any confusion.} must
admit a description in terms of the two dimensional CFT that is
dual to the D1/D5 system.

For certain values of the parameters the decoupled metric is
everywhere locally $AdS_3 \times S^3$, for instance for $R=0$ the
solution reduces to the BMPV black hole
\cite{Breckenridge:1996is}. In \cite{Elvang:2004ds} (see also
\cite{Emparan:2005jk}) the near horizon limit of the solution was
studied in detail, where it was shown that provided the following
condition is satisfied
\begin{equation}
\label{irfactor} {\cal Q}_1 q_1+{\cal Q}_2 q_2-{\cal Q}_3 q_3= 2
q_1 q_2 m R_z
\end{equation}
for some integer $m$, the space factorizes into the near horizon
limit of the extremal BTZ black hole times the quotient space
$S^3/{\mathbb Z}_{q_3}$. In this near-horizon limit the black ring
is no longer visible, so we will restrict our attention to the
first decoupling limit only.

\subsection{Decomposition of the fluctuations and vev's}

In the previous section we have seen that the decoupling limit of
the three charges supertube yields a space that is asymptotically
$AdS_3 \times S^3$. To extract the one-point functions of CFT
operators, we need to decompose the full solution into those
degrees of freedom that diagonalize the linearized equations of
motion; the leading large $r$ behavior of each of those degrees of
freedom then provides us with the one-point functions. To obtain
the properly normalized one-point functions we also need to know
the precise normalization with which each degree of freedom
appears in the action, after we expand to action to second order.
Luckily, this analysis was already done in
\cite{Deger:1998nm,Arutyunov:2000by}.

In order to diagonalize the linearized equations of motion, and
also in order to extract the quantum numbers of the dual CFT
operators, it is useful to organize all fluctuations in terms of
representations of the isometry group $SO(2,2)\times SO(4)$ of
AdS${}_3\times$S${}^3$. At the linearized level, different
representations cannot mix with each other, and must therefore
couple to different operators in the CFT. This works as long as we
perturb around global AdS${}_3\times$S${}^3$. However, in view of
(\ref{zeroorder}), it might be more natural to think of the black
ring as ``small'' perturbation of the $M=0$ BTZ solution rather
than as a ``large'' perturbation around global
AdS${}_3\times$S${}^3$. Because the $M=0$ BTZ solution is locally
the same as AdS${}_3$, it will have locally the same Killing
vector fields as AdS${}_3$, but these will not be globally
well-defined; their explicit expressions are summarized in
appendix~A.1. Though it would be interesting to develop this point
of view in more detail, we will in this paper view the black ring
as a perturbation of global AdS${}_3$, which is the same as the
$M=-1$ BTZ black hole. The relevant $SO(2,2)\times SO(4)$
generators are given in appendix~A.2 and~B. The $SO(2,2)\simeq
SL(2,\mathbb R) \times SL(2,\mathbb R)$ generators, when viewed as
vector fields on the $M=0$ BTZ describe globally well-defined
vector fields, but they generate only asymptotic isometries, not
global ones. Eventually, all these differences are more or less
irrelevant, since the one-point functions are obtained from the
leading asymptotics of the solutions of the field equations only.

We now turn to the analysis of the black ring solution. We write
the full metric and three form in the following way
\begin{equation}
g_{\mu \nu}=g_{\mu \nu}^{(0)} + h_{\mu \nu},\hspace{0.3in}
F=F^{(0)}+ d b
\end{equation}
Then, following \cite{Deger:1998nm,Arutyunov:2000by} we
parametrize the metric fluctuations around the $AdS_3 \times S^3$
background as (we restrict ourselves to the sector describing
scalar fields on AdS${}_3$)
\begin{equation}
h_{ab}=h_{(ab)}+\frac{1}{3}g^{(0)}_{a b} h^c_c,\hspace{0.2in}
(g^{(0)})^{ab}h_{(ab)}=0
\end{equation}
and we write the two form potential (which only has a self dual
part) in terms of a vector field $U^c$ as:
\begin{equation}
b_{ab}= 2 \sin\theta \cos\theta \epsilon_{abc}U^c
\end{equation}
with $a,b=1,2,3$ $SO(3)$ vector indices (i.e. $S^3$ tangent space
indices) which are raised and lowered with the $S^3$ metric
$g^{(0)}_{ab}$.

We can then expand the linearized fluctuations in terms of
harmonic functions on $S^3$ (see appendix~B for a detailed
discussion on harmonics on $S^3$ )
\begin{eqnarray}
h^a_a(r,\theta) & = & \sum_{k=0}^{\infty}\pi^{k}(r) Y^{k}_s(\theta)\\
h_{ab}(r,\theta) & = & \sum_{k=4}^{\infty}\varrho^{\pm}_k(r) (Y_{t
\pm}^k)_{a b} + \sum_{k=2}^{\infty} \zeta^{k}(r) \nabla_{(a}
\nabla_{b)} Y^{k}_s(\theta)\\
U_a (r,\theta) & = & \sum_{k=2}^{\infty} U^{k}(r) \partial_a
Y^{k}_s(\theta).
\end{eqnarray}
By using the completeness relations presented in the appendix one
can easily check that the component $\zeta^{k}(r)$ is different
from 0 for the solution under consideration, that means we are
outside the de Donder gauge. This component can be easily removed
(as we are interested only in the leading $1/r$ piece) by an
appropriate gauge transformation
\be
\delta h_{ab}=\nabla_a \xi_b + \nabla_b \xi_a, \qquad \delta U_a =
\xi_a
\ee
with
\begin{eqnarray}
\xi_a&=&\delta_{a \theta} \sum_{i=1}^{\infty} d_{2i}
\frac{R^{2i}}{r^{2i}}\partial_\theta Y^{2i}_s(\theta)\\
d_i&=&-(-1)^{i/2}\frac{\sqrt{i+1}(i-2)!}{(i+2)!}\frac{1}{C_{i/2}}
\left(1-\frac{q_3^2 R^2}{Q_1 Q_2} \right)
\end{eqnarray}
where we have restricted ourselves to the regime of large charges,
more precisely we rescale $Q_1 \rightarrow Q_1/\epsilon,Q_2
\rightarrow Q_2/\epsilon,R \rightarrow R/\epsilon$ and then take
$\epsilon \rightarrow 0$ and keep the leading term in the
$\epsilon$-expansion; $C_{k}=\frac{2k!}{(k+1)!k!}$ denote the
Catalan numbers. We now obtain
\begin{eqnarray}
\varrho^{\pm}(r)&=&(-1)^{k/2}\frac{k \sqrt{k-2}
}{(k(k-1))^{3/2}}\frac{1}{C_{k/2-1}}\frac{R^k}{r^k}\left(1-\frac{q_3^2
R^2}{Q_1 Q_2} \right)+\ldots\\
\pi^{k}(r)&=&\frac{3(-1)^{(k/2)}k}{2(k-1)^2\sqrt{k+1}}\frac{1}{C_{k/2-1}}
\frac{R^k}{r^k}\left(1-\frac{q_3^2 R^2}{Q_1 Q_2} \right)+\ldots \\
U^{k}(r)&=&-\frac{(-1)^{(k/2)}}{4(k-1)^2\sqrt{k+1}}\frac{1}{C_{k/2-1}}
\frac{R^k}{r^k}\left(1-\frac{q_3^2 R^2}{Q_1 Q_2} \right)+\ldots
\end{eqnarray}
where the dots denote terms that are subleading in $1/r$ or in the
large charge limit explained above.\footnote{Using the results in
appendix~B it is straightforward to compute the full dependence on
the charges (at leading order in the $1/r$ expansion) but we could
not find a manageable closed expression for general level $k$.}
Note that the value for $\varrho^{\pm}(r)$ does not depend on the
gauge choice. We have also rescaled the full six-dimensional
metric by a factor of $1/\sqrt{Q_1 Q_2}$ so that the $S^3$ has
unit radius (as done for example in  \cite{Mihailescu:1999cj}).
This results in a prefactor in the six-dimensional supergravity
actions which equals, up to factors of $2\pi$, the product of the
inverse ten-dimensional Newton constant, the volume of the
four-torus and $Q_1Q_2$. We will denote this number by $N$, since
it is equal to
\be
N\equiv \frac{1}{g_2^2 \ell_s^8} \ell^4 Q_1 Q_2 = N_{D1} N_{D5}.
\ee
In order to diagonalize the linearized equations of motion we need
to make the following field redefinition
\begin{eqnarray}
\pi^{k}=-6k
\sigma_k+6(k+2)\tau_k,\hspace{0.2in}U^k=\sigma_k+\tau_k
\end{eqnarray}
Then, at leading order in the $1/r$ expansion (as we will see the
vanishing of $\tau$ is consistent with its conformal dimension)
\begin{eqnarray}
\sigma_k&=&-\frac{(-1)^{(k/2)}}{4(k-1)^2\sqrt{k+1}}\frac{1}{C_{k/2-1}}
\frac{R^k}{r^k}\left(1-\frac{q_3^2 R^2}{N} \right)+\ldots\\
\tau_k &\approx& 0 \label{ww2}
\end{eqnarray}
To extract the one-point functions from the large-$r$ behavior we
first consider the action for a general massive scalar field
$\phi$
\begin{eqnarray}
S=\frac{\eta_{\phi}}{2}\int_{AdS_{d+1}}\sqrt{G^{0}}\left(-\nabla_\mu
\phi \nabla^\mu \phi-m_{\phi}^2 \phi^2 \right) \label{w1}
\end{eqnarray}
The scalar field $\phi$ will act as a source of a dual boundary
operator ${\cal O}_\phi$ and the two point function of this
operator can be computed to be \cite{Freedman:1998tz}
\begin{eqnarray}
< {\cal O}(x) {\cal
O}(y)>=\eta_{\phi}\frac{\Gamma[\Delta+1]}{\pi^{d/2}\Gamma[\Delta-d/2]}
\frac{2\Delta-d}{\Delta}\frac{1}{|x-y|^{2\Delta}}
\end{eqnarray}
where $\Delta=\frac{1}{2}(d + \sqrt{d^2+4m^2})$, for our
particular case we should obviously set $d=2$. The decoupled black
ring is asymptotically AdS${}_3\times$S${}^3$ and therefore
corresponds to a normalizable deformation, described by
expectation values for all operators ${\cal O}$. If the solution
of (\ref{w1}) near the boundary behaves as
\begin{equation}
\phi \rightarrow 1/r^{\Delta}(A(x)+...) \hspace{0.3in} \mbox{for}
~ r \rightarrow \infty
\end{equation}
then the expectation value for the operator ${\cal O}$ is
\cite{Klebanov:1999tb}
\begin{equation} \label{ww1}
<{\cal O}(x)>=\eta_\phi(2\Delta-d) (Q_1 Q_2)^{-\Delta/2}A(x) .
\end{equation}
The extra factor $\eta_\phi$ was put to one in the conventions of
\cite{Klebanov:1999tb}, but it is relevant for us since it
contains a non trivial dependence on $N$. The extra factor $(Q_1
Q_2)^{-\Delta/2}$ comes from the fact that \cite{Arutyunov:2000by}
and \cite{Klebanov:1999tb} use a different coordinate system with
a radial coordinate $\rho$ in which the boundary is at $\rho=0$.
The coordinate $\rho$ is related to our coordinate $r$ by
$r=\frac{\sqrt{Q_1 Q_2}}{\rho}$, and the change of variables
yields the extra factor in (\ref{ww1}).

From \cite{Arutyunov:2000by} we can read off
\begin{eqnarray}
\eta_{\sigma^k} = \frac{N}{(2\pi)^3}16k(k-1),\hspace{0.2in}\eta_{\varrho^k} = \frac{N}{2(2\pi)^3}\\
m_{\sigma^k}^2= m_{\varrho^k}^2 =k(k-2) \Rightarrow
\Delta_{\sigma^k}=\Delta_{\varrho^k} =k,\\ m_{\tau^k}^2=(k+2)(k+4)
\Rightarrow \Delta_{\tau^k}=k+4
\end{eqnarray}
Notice that the values of $\Delta$ for each scalar field are in
correspondence with the leading behavior of the field near the
boundary, in particular they imply the vanishing of $\tau^k$ at
order $1/r^k$, which is consistent with (\ref{ww2}).

For the fields under consideration
\begin{eqnarray}
< {\cal O}_{\sigma^k} (x) {\cal O}_{\sigma^k} (y)>=\frac{4 N}{\pi^4}k(k-1)^3 \frac{1}{|x-y|^{2k}}\\
< {\cal O}_{\varrho^k} (x) {\cal O}_{\varrho^k}
(y)>=\frac{N}{9\pi^4}(k-1)^2 \frac{1}{|x-y|^{2k}}.
\end{eqnarray}
If we normalize operators in such a way that the two point
correlation function is $1/|x-y|^\Delta$ we finally obtain
\begin{eqnarray}
<{\cal O}_{\sigma^k}>&=& -\frac{2(-1)^{k/2}}{\pi} N^{1/2}
\frac{k}{(k+2)\sqrt{k(k^2-1)}} \frac{1}{C_{k/2}}
\left(\frac{R}{\sqrt{Q_1Q_2}}\right)^k \left(1-\frac{q_3^2 R^2}{N} \right)\\
< {\cal O}_{\varrho^k}>&=& \frac{3(-1)^{k/2}}{8 \pi} N^{1/2}
\frac{k \sqrt{k-2}}{(k(k-1))^{3/2}} \frac{1}{C_{k/2-1}}
\left(\frac{R}{\sqrt{Q_1 Q_2}}\right)^k \left(1-\frac{q_3^2
R^2}{N} \right).
\end{eqnarray}
This can be rewritten in a straightforward way in terms of the
integer charges, using (\ref{integercharge}).

\section{1/2 BPS case}

A particularly interesting case of the above solution is the so
called "small black ring"
\cite{Iizuka:2005uv,Balasubramanian:2005qu} obtained from the
general solution by setting \footnote{Actually in this limit the
solution is singular but it is believed to become a small black
ring with a string scale horizon once higher order curvature
corrections are taken into account.}
\begin{equation} \label{smallbr}
q_1=q_2=Q_3=0
\end{equation}
In this case (\ref{noccc}) is trivially saturated but the absence
of closed causal curves still imposes the non-trivial constraint
\begin{equation} \label{ccc2}
q_3 R^2 \leq \frac{Q_1 Q_2}{q_3}.
\end{equation}
From now on we will express everything in terms of the integer
charges using (\ref{integercharge}). We will also put $R_z=1$, as
this is the radius of the circle of the cylinder on which the dual
CFT lives, so the $R_z$ dependence can always be restored by a
conformal transformation. Then $q_3=n_{KK}$ and we will use $q_3$
instead of $n_{KK}$ to denote this particular integer. The absence
of closed causal curves (\ref{ccc2}) becomes simply \be q_3 J \leq
N . \ee One can easily check that the macroscopic entropy vanishes
in the limit (\ref{smallbr}), however the system still has a
finite microscopic entropy
\cite{Russo:1994ev,Iizuka:2005uv,dabholkarrecent}
\begin{equation}
\label{microent} S_{micro}=4 \pi \sqrt{N-q_3 J}
\end{equation}
When considering the system compactified on K3 evidence was given
in \cite{Iizuka:2005uv,{dabholkarrecent}} that the solution
develops a non-vanishing horizon once stringy $R^2$ corrections to
the supergravity action are considered, furthermore, the area of
such a horizon gives rise to the entropy (\ref{microent}). The
full geometry including $R^2$ corrections, however, has not been
computed.

From (\ref{fullarea}) we see that a way to get the small black
ring microscopic entropy is by setting\footnote{Notice that $Q_3$
is such that (\ref{irfactor}) is satisfied with $m=0$, but in fact
in the limit of large charges we will reproduce the entropy for
any finite $m$. With the values $n_{D1}=n_{D5}=1$ the entropy will
have the correct form but with a different proportionality factor,
in general the black ring entropy reduces to $S_{micro}=2 \pi
\sqrt{n_{D1}n_{D5}} \sqrt{N-q_3 J}$, so we must set e.g.
$n_{D1}=n_{D5}=2$ in order to reproduce the microscopic entropy
with the correct proportionality factor for the system on K3, and
e.g. $n_{D1}=1$ and $n_{D5}=2$ for the system on T${}^4$. }
\begin{eqnarray}
\label{smallplusR}
n_{D1}=n_{D5}=1  \nonumber \\
Q_3=\frac{q_1 Q_1+q_2 Q_2-q_1 q_2 q_3}{q_3}.
\end{eqnarray}
Notice that the entropy can by written in the following
alternative form \begin{equation} S=2\pi \sqrt{[N_1 N_2-\bar{N}_1
\bar{N}_2]N_3+ [\bar{N}_1 \bar{N}_2-n_3 J ] n_1 n_2- J_{\phi}^2},
\end{equation} and as argued in \cite{Elvang:2004ds,Bena:2004tk} this suggests the
interpretation that the system decomposes into two sectors, the
"BMPV" sector, with central charge $c'=6 [N_1 N_2-\bar{N}_1
\bar{N}_2]$ and angular momentum $J_{\phi}$ and the "supertube"
sector, with central charge $c''=\bar{N}_1 \bar{N}_2$ and angular
momentum $J$. The condition (\ref{smallplusR}) implies $c' N_3=6
J_{\phi}^2$ so that the BMPV sector carries no entropy. It would
be interesting to understand better whether there is a physical
mechanism behind the decoupling of the BPMV sector. In any case we
would like to conjecture that once we include $R^2$ corrections to
the small black ring, the geometry we get is similar to the
general black ring with the conditions (\ref{smallplusR}).

The small black ring limit is particularly interesting since the
bulk geometry is $1/2$-BPS, this implies that typical states that
describe it will also be 1/2 BPS and the corresponding density
matrix in the dual CFT  is of the form $\sum_{a,b} \rho_{ab}
|a\rangle\langle b|$, with $|a\rangle$ and $|b\rangle$ chiral
primaries, as we will see in the next section, this will simplify
the computation of the one-point functions in the orbifold CFT.

It is easy to see that in the limit of large charges considered in
the previous section the answer for the one point functions
obtained from supergravity in the small black ring limit are
independent on whether we choose $q_1=q_2=Q_3=0$ or
(\ref{smallplusR}). Expressed in terms of $N=N_{D1}N_{D5}$, $J$
and $q_3$ dependence, we obtain
\begin{equation} \label{finalsugra}
< {\cal O} > \sim  N^{1/2} q_3^{-k/2} \left( \frac{J}{N}
\right)^{k/2} (1 -q_3 \frac{J}{N} ) .
\end{equation}
Our aim will be to obtain the same result from the dual orbifold
CFT.

\bigskip

In \cite{Bena:2004tk} a microscopic interpretation of the small
black ring was given (see also \cite{Iizuka:2005uv} for the case
$q_3=1$), in the regime $J \sim N$ the correct entropy and angular
momentum are accounted for if we consider states of the form
\begin{equation} \label{ww4}
(a^{+}_{n=-q_3})^J \times \prod_{n=1}^{\infty}
\left(\prod_{i=\pm,3,...,24} (\alpha_{-n}^i)^{N_{n_i}} \right)|0>
\end{equation}
where the first factor represents a Bose-Einstein condensate and
accounts for the angular momentum, and the second factor has no
net angular momentum and accounts for the entropy. The notation
will be further explained in the next section but is based on the
identification of the space of chiral primaries with the states at
level $N$ in a Fock space built out of 24 free bosons. In section
\ref{toymodel} we present a thermodynamic toy model which will
lead to the Bose condensate in (\ref{ww4}) for general $q_3$.

\section{Orbifold computation}

\subsection{Preliminaries}

In this section we would like to compare the one-point functions
obtained from supergravity to those obtained from the dual
conformal field theory. As is well-known, the dual conformal field
theory is believed to be a deformation of the $N=(4,4)$ orbifold
SCFT which is a sigma model with target space $M^N/S_N$, with
$M=T^4$ or K3 (see e.g.
\cite{{Strominger:1996sh},{Seiberg:1999xz},{deBoer:1998ip},{Maldacena:1998bw},
{deBoer:1998us},{Dijkgraaf:1998gf},{Larsen:1999uk},{David:1999ec}}.
We will perform our calculations at the orbifold point. In the
1/2-BPS case one may hope that the results will be independent of
the deformation, though we are not aware of any proof of this; in
general the results will certainly depend on it.

Since we will in this section focus on the $1/2$ BPS case, we will
consider density matrices of the form $\sum_{a,b} \rho_{ab}
|a\rangle\langle b|$ with $|a\rangle$ and $|b\rangle$ chiral
primaries. The set of chiral primaries is most easily understood
by dualizing the D1-D5 system on $M$ to a F1-P system in type II
(for $M=T^4$) or in the heterotic $E_8\times E_8$ string (for
$M=K3$). Then we find that the space of chiral primaries, as a
vector space, is the same as the set of states at level $N$ in a
Fock space built out of $b^{\rm even}=b^0+b^2+b^4$ bosonic
oscillators and $b^{\rm odd}=b^1+b^3$ fermionic oscillators, with
$b^i = {\rm dim}\,H^i(M)$ the Betti numbers of $M$. Of course the
chiral primaries also form a ring, but this ring structure is much
more difficult to obtain from the F1-P system
\cite{Harvey:1995fq,Harvey:1996gc}. The ring structure is very
interesting and has been the subject of various mathematics papers
in the past few years (see e.g. \cite{lehnsorger,ruan,wang}). It turns out
to be highly non-trivial to express the ring structure in terms of
the free bosons and fermions and we will not discuss this
here\footnote{There are many interesting questions: is the ring
structure related to an integrable system, to $W$-algebras, does
it become something simple at large $N$, is it related to matrix
models of some sort, is there a collective field theory
description at large $N$ \cite{Jevicki:1998bm}, etc.} In any
case, most one-point functions require information which is not
contained in the chiral ring.

Because for $M=K3$ all elements of the chiral ring are bosonic, we
will only discuss this case here. It also has the advantage that
for $M=K3$ one can explicitly include $\alpha'$ corrections in the
supergravity solutions and see a horizon form in case the
classical horizon area vanishes \cite{Dabholkar:2004yr}. Thus this
is a natural case in which to study various aspects of black hole
physics.

The free boson description of the chiral ring can also be
extracted directly from the orbifold CFT. The latter  has various
twisted sectors labeled by conjugacy classes of $S_N$. Any group
element of $S_N$ can be written in its cycle decomposition as
$(c_1)(c_2)\ldots$ with $|c_i|$ the length of the cycles. Each
cycle $(c_i)$ describes $i$ copies of $M$ that are being
cyclically twisted as we move once around the string. In other
words each cycle describes effectively a long string of length
$|c_i|$. Each such long string gives rise to a set of chiral
primaries, one for each element $\gamma\in H^{\ast}(M)$. In the
twisted sector $(c_1)(c_2)\ldots$ we therefore find operators that
we will denote as $\sigma_{c_1}(\gamma_1) \sigma_{c_2}(\gamma_2)
\sigma_{c_3}(\gamma_3)\ldots$. However, to get an honest operator
in the orbifold CFT we need to sum over the centralizer of the
group element $(c_1)(c_2)\ldots$, and also over the entire
conjugacy class this group element belongs to. Notice that
$\sigma_{c}(\gamma)$ refers to a particular cyclic permutation $c$
and not yet to its conjugacy class. We will denote the operator
obtained by averaging $\sigma_{c}(\gamma)$ over its conjugacy
class by $\sigma_{|c|}(\gamma)$. This is an honest operator in the
orbifold CFT, and corresponds to the bosonic creation operator
$\alpha_{-|c|}^{\gamma}$ in the free boson description.

It is important to keep in mind that the free boson representation
does not reflect the chiral ring structure. The product of the two
operators $\sigma_{|c_1|}(\gamma_1)$ and
$\sigma_{|c_2|}(\gamma_2)$ is quite complicated. Each operator
involves a sum over conjugacy classes, and therefore in the
product there will be terms where the two cycles are disjoint, but
also where the two-cycles overlap. Taking the naive product of the
boson creation operators would miss the second type of
contributions.

Chiral primaries have (in the NS sector) conformal weights
$(h_L,h_R)$, with $J=h_L-h_R$ the angular momentum. These
conformal weights are related to the Hodge decomposition of the
complex four-manifold $M$. A form $\gamma$ of degree $(p,q)$
yields a chiral primary with weights $(h_L,h_R)=(p/2,q/2)$,
whereas $\sigma_{|c|}(\gamma)$ yields a chiral primary with
weights $(h_L,h_R)=((p+|c|-1)/2,(q+|c|-1)/2)$ in the orbifold
theory. The cohomology of K3 has unique elements of degrees
 $(0,0)$, $(2,0)$,
$(0,2)$ and $(2,2)$. The corresponding operators
$\sigma_{|c|}(\gamma)$ are denoted by $\sigma_n^{--}$,
$\sigma_n^{+-}$, $\sigma_n^{-+}$ and $\sigma_n^{++}$ with $n=|c|$
in \cite{Lunin:2001pw}. These operators exist for any
hyperk\"ahler $M$ and can be constructed using just the $N=4$ SCFT
without explicit reference to $M$. For $M=K3$ there are also 20
elements of degree $(1,1)$. Therefore, the chiral primaries of
$K3^N/S_N$ can be represented in terms of the Fock space
constructed out of 24 bosons. Of these, 22 carry no angular
momentum, one carries angular momentum $+1$ and one $-1$.

\subsection{The sum over conjugacy classes}

The solutions of the field equations of 6d supergravity on
AdS${}_3\times$S${}^3$ (obtained by KK reduction of type II string
theory on K3) can be put in one-to-one correspondence with the
``single particle'' chiral primaries $\sigma_{|c|}(\gamma)$ and
their superconformal descendants, at least as far as the quantum
numbers go \cite{{Larsen:1998xm},{deBoer:1998ip}}. Strictly speaking it
has not been shown that the supergravity fields are dual to
precisely these one-particle operators and that they are not
modified by products of operators with smaller ${|c|}$. We will
nevertheless use this as our working assumption. A good test would
for example be to compare orbifold three-point functions to the
dual supergravity result. A detailed comparison has not been done
but see \cite{{Lunin:2001pw},{Mihailescu:1999cj}}. It is also
plausible that at large $N$, the leading contribution simply comes
from the single particle operator $\sigma_{|c|}(\gamma)$ and its
superconformal descendants. A simple estimate shows that the
admixture of operators based on multiple cycles is suppressed by
factors of $N^{-1/2}$. As we will only focus on the leading large
$N$ behavior of correlation functions, this puts further faith in
our working assumption.

A further simplifying assumption that we will be making is that
the density matrix in which we compute one-point functions is
diagonal in the twist field basis. This seems to include all
examples of interest, and once we drop this assumption the
calculation become quickly untractable.

With these simplifications the correlations functions that we need
to compute are sums of correlators of the form
 $\langle A | {\cal O}| A
\rangle$ with
\begin{equation} {\cal O}= (J_-)^u(\bar{J}_-)^{\bar{u}} \sigma_{|c|}(\gamma)
=(J_-)^u(\bar{J}_-)^{\bar{u}} \sum_{{\rm cent\,\,conj}} \sigma_{c}(\gamma)\ee and $A$ a chiral
primary of the form \be A= \sum_{{\rm cent\,\,conj}} \prod_i
\sigma_{c_i}(\gamma_i)
\end{equation}
where $c_i$ are disjoint cycles and $\sum_{{\rm cent\,\,conj}}$
indicates a suitably normalized sum over the centralizer and
conjugacy class of the corresponding group element. The
supergravity operators that will have a nonzero vev should clearly
have $U(1)_L\times U(1)_R$ charge equal to zero. Therefore ${\cal
O}$ cannot be a chiral primary, it has to be a descendant of a
chiral primary in such a way that the $U(1)_L\times U(1)_R$ charge
vanishes. To achieve this we lower the charges of ${\cal O}$ using
the $SU(2)_L \times SU(2)_R $ lowering operators. There are also
other descendants that make use of the supersymmetry generators
that could have a one-point vev. Our calculations and estimates
below apply to such operators as well.

The sums over centralizers and conjugacy classes are important for
the overall normalization and the large $N$ dependence of the
various correlators. The centralizer of a group element
$(c_1)(c_2)\ldots$ consists of those group elements that either
cyclically permute the cycles, or that permute cycles of equal
length. If we assume that there are $r_i$ cycles of length $i$,
then the order of the centralizer is  \be z[\{c\}]= \prod_i
i^{r_i} r_i!. \label{dimcent} \ee Once we associate different
classes $\gamma_i\in H^{\ast}(M)$ to the cycles $c_i$, the
centralizer of the group element no longer acts trivially on the
state. A sum over the centralizer effectively symmetrizes the
classes $\gamma_i$ over the different cycles of equal length.
Incidentally, this is also why the corresponding operators
$\alpha_{-|c|}(\gamma)$ obey Bose statistics. It will be
convenient in what follows to separate the centralizer in a piece
which preserves the state, and a remainder. The piece that
preserves the state consists again of the cyclic permutations of
the cycles, plus permutations of those cycles of equal length that
carry the same label $\gamma_i$. The remaining elements of the
centralizer together with the sum over conjugacy classes form a
group $G(A)$. We will explicitly keep track of this group in what
follows. Altogether the precise definition of our operator $A$ now
reads \be A=\frac{1}{N(A)} \sum_{g\in G(A)} g(\prod_i
\sigma_{c_i}(\gamma_i)), \ee where the normalization constant
$N(A)$ will be chosen such that $\langle A|A\rangle=1$\footnote{As
usual, we assume here that
  $|A\rangle=A(0)|0\rangle$ and that $\langle A|$ is the BPZ conjugate
  of $|A\rangle$.}. In addition, we will assume the operators
$\sigma_{c}(\gamma)$ are normalized so that $\langle
\sigma_{c}(\gamma) | \sigma_c(\gamma) \rangle=1$. In general, if
we have operators $A_i$ living in a sector twisted by $g_i\in
S_N$, the correlator of the $A_i$ on the sphere will vanish unless
$\prod_i g_i=1$. If $|A\rangle$ is twisted by $g\in S_N$, then
$\langle A|$ is twisted by $g^{-1}$, and when we compute $\langle
A|A\rangle$ only terms with $g=g'$ contribute to $N(A)^{-2}
\sum_{g\in G(A)} \sum_{g'\in G(A)}$. Therefore, \be \langle
A|A\rangle = \frac{|G(A)|}{N(A)^2}. \ee The order of the group
$G(A)$ can be expressed as follows. Suppose that there are
$r_i(\gamma_j)$ occurrences of operators $\sigma_{c_l}(\gamma_j)$
with $|c_l|=i$ in $A$. Then by a simple generalization of
(\ref{dimcent}) we find that \be |G(A)| = \frac{N!}{\prod_{i,j}
i^{r_i(\alpha_j)} r_i(\alpha_j)! } \ee and it is clear that the
normalization factor will be \be N(A)=\sqrt{|G(A)|}. \ee

The ``single-particle'' operator ${\cal O}$ is part of the
$g$-twisted sector of the theory where $g$ consists of just one
cycle $c$ with length $|c|$, which we will denote by $k$ from now
on. An analysis similar to the one we just did shows that in order
to properly normalize ${\cal O}$ we need an extra factor of \be
\frac{1}{N({\cal O})} = \frac{1}{\sqrt{ |G({\cal O})| }} = \left(
\frac{ (N-k)! k }{N!} \right)^{\frac{1}{2}}. \ee

The final correlator therefore involves a sum \be \sum_{g\in G(A)}
\sum_{g'\in G(A)} \sum_{g''\in G({\cal O})} \frac{1}{N(A)^2 N(O)}
\ee and there can only be nonzero contributions if the three group
elements are suitably aligned. More precisely, denote by $g_A$ and
$g_{\cal O}$ the twists in $A$ and ${\cal O}$ before doing the
sum. Then the only non-vanishing contributions can arise when \be
\label{relg} (g g_A g^{-1}) (g'' g_{\cal O} (g'')^{-1})  = g' g_A
(g')^{-1}. \ee To make a precise counting of the number of sets of
three group elements for which this holds is a daunting task.
Since $g_A$ is a relatively ``long'' group element, and $g_{\cal
O}$ is a relatively ``short'' group element, one expects that the
single cycle of length $k$ in $g_{\cal O}$ will only interact with
a few of the cycles in $g_A$. Geometrically, this means the
following. The three group elements $g_1=g g_A g^{-1}$, $g_2=g''
g_{\cal O} (g'')^{-1}$ and $g_3=g' g_A (g')^{-1}$ obeying
(\ref{relg}) define an $N$-fold cover $\Sigma\subset M^N$ of the
string world-sheet $\mathbb P^1$, such that the map $\Sigma\rightarrow
\mathbb P^1$ has three branch points with monodromies $g_1$, $g_2$
and $g_3$ at $0,1,\infty$. In general, $\Sigma$ will be
disconnected; Since $g_2$ consists of a single cycle, there must
be a single connected component $\Sigma_0$ of $\Sigma$ that
contains a branch point with monodromy $g_2$. This distinguished
connected component is also a branched cover of $\mathbb P^1$ and
its degree will be denoted by $s$. As ${\cal O}$ does not
contribute to the remaining components of $\Sigma$, the
calculations involving these other components reduce to that of
two-point functions, and they can be done exactly as we did above
with the same combinatorics. Suppose that of all the cycles in
$A$, $d_i(\gamma_j)$ cycles of length $i$ with operator $\alpha_j$
inserted ``live'' on $\Sigma_0$, by which we mean that the
monodromies of $\Sigma_0\rightarrow \mathbb P^1$ at $0,\infty$ are
group elements consisting of $d_i(\gamma_j)$ cycles of length $i$.
These cycles somehow interact with the single cycle of length $k$
over in total $s$ elements in an irreducible way. Clearly,
$\sum_{i,j} i d_i(\gamma_j)=s$. There can in general be different
topological ways to do this, but one would expect the number of
topologically inequivalent contractions (i.e. ones that are not
related by overall conjugation) to be very small. Call this number
$T$. In addition, there could be a few overall conjugations that
leave the configuration precisely invariant. Again, this will be a
small number, perhaps one can even prove it will be always one.
Denote this number by $R$.

The additional combinatoric factors that appear are the number of
conjugacy classes of $A$ for those cycles which does not belong to
$\Sigma_0$; this equals \be L= \frac{(N-s)!}{\prod_{i,j}
i^{r_i(\gamma_j)-d_i(\gamma_j)} (r_i(\gamma_j)-d_i(\gamma_j))!} .
\ee Next, there are $s!$ different ways to do an overall
conjugation of the three monodromies of $\Sigma_0$. Accidental
symmetries and different topological contraction possibilities are
taken into account by the factor $T/R$. Finally, there are $\left(
\begin{array}{c}  N \\ s
\end{array} \right)$ different possible ways to choose the
irreducible component $\Sigma_0$ in $M^N$. The final total
combinatorial prefactor is therefore \be K=
 \frac{1}{N(A)^2 N(O)}\left( \begin{array}{c}  N \\ s  \end{array} \right)
s! L \frac{T}{R}. \ee Massaging this a bit we obtain \be K=\left(
\frac{ (N-k)! k }{N!} \right)^{\frac{1}{2}} \frac{T}{R}
\prod_{i,j} \frac{ i^{d_i(\gamma_j)}
r_i(\gamma_j)!}{(r_i(\gamma_j)-d_i(\gamma_j))!}. \ee

\subsection{Large $N$ analysis}

The combinatorial factor $K$ controls most of the $N$ dependence of the
correlation function, since all that remains is a relatively small
correlator times this combinatorial factor. To study large $N$, it
is useful to use the following observation. In the irreducible
component $\Sigma_0$ of degree $s$, there are three branch points
with monodromy $h_1$, $h_2$ and $h_3$ whose product equals one.
Suppose that the number of cycles of each of these group elements,
viewed as elements of $S_s$, is $y_i$. Then the genus of
$\Sigma_0$ is \be \label{genus} g=\frac{1}{2} ( s+2-y_1-y_2-y_3)
\ee which clearly has to be a nonnegative integer. This is called
the graph defect in \cite{lehnsorger}, since we can associate a
simple graph to the three group elements and use the combinatorial
genus of the graph to obtain (\ref{genus}). The number of orbits
in the piece from $A$ that lives on $\Sigma_0$ is \be
y_1=y_3=\sum_{i,j} d_i(\gamma_j). \ee The number of orbits from
${\cal O}$ is $s-k+1$. Thus \be g=\frac{1}{2} (k+1 -2 \sum_{i,j}
d_i(\gamma_j) ). \ee For large $N$ and finite $k$ $N!/(N-k)! \sim
N^k$. Thus the combinatorial factor behaves for large $N$ as
\be
\label{comb} N^{\frac{1}{2} - g -\sum_{i,j} d_i(\gamma_j) }
 \prod_{i,j} \frac{
i^{d_i(\gamma_j)} r_i(\gamma_j)!}{(r_i(\gamma_j)-d_i(\gamma_j))!}.
\ee As a check, if $A$ consists of a single cycle, we find that
the correlation function scales as $N^{-1/2}$, which is the
correct answer. Similarly, if $A$ is a relatively simple operator,
it is easy to extract the large $N$ dependence. However, we are
mainly interested in the case where $A$ is very complicated and
consists of many cycles which are typically randomly distributed.
In such a situation, there are many different contractions which
contribute to the correlation function. The combinatorial factor
involves (for $r_i(\gamma_j) \gg d_i(\gamma_j)$) a factor \be
\label{aux12} \prod_{i,j} (i r_i(\gamma_j))^{d_i(\gamma_j)} . \ee
This still has to be multiplied by the appropriate correlation
function, summed over all possible sets of choices of the numbers
$d_i(\gamma_j)$, and averaged over the ensemble to which the state
$|A\rangle$ belongs. If we ignore the contribution from the
correlation function, and assume that the total length of the set
of cycles from which we randomly select $d_i(\gamma_j)$ of type
$\sigma_i(\gamma_j)$ is $P$, then the factor (\ref{aux12}) appears
to scale like $P^{\sum_{i,j} d_i(\alpha_j)}$. For example, if
$P=N$, then (\ref{aux12}) appears to scale like $N^{\sum_{i,j}
d_i(\alpha_j)}$ which would cancel against a similar factor in
(\ref{comb}). In other words, it would appear that arbitrarily
complicated contractions with genus $g=0$ would contribute equally
at large $N$. This is a very peculiar conclusion and there are
several reasons why we believe it is incorrect. First of all, we
ignored the contributions of the actual correlation functions.
These are notoriously difficult to compute, but will certainly be
nontrivial functions of $i$ and $j$ (see e.g. \cite{Lunin:2001pw}
for a three-point function calculation) and putting these in may
well change this naive conclusion. Furthermore, it is quite
possible that averaging over an ensemble will involve various
signs that suppress the above naive estimate. Finally, if the
above estimate were correct, it would predict that in the case of
the $M=0$ BTZ black hole all one-point functions would be turned
on with equal strength. This is certainly not the case. To leading
order in the $\alpha'$ expansion, only untwisted operators in the
$M=0$ BTZ have a nonzero one-point function. This situation will
probably change once higher order corrections are included (after
all the $M=0$ BTZ black hole has a residual entropy of order
$\sqrt{N}$), but those higher order corrections will be suppressed
by higher orders of $\alpha'$, i.e. $N^{-1/2}$. It would be
interesting to work this out in more detail (see also
\cite{Dabholkar:2004yr}), but for us it motivates us to conjecture
that after including correlation functions, and after averaging
over a random ensemble of cycles of total length $P$, the factor
(\ref{aux12}) scales as $P^{(\sum_{i,j} d_i(\gamma_j) +1)/2}$: \be
\label{aux13} \prod_{i,j} (i r_i(\gamma_j))^{d_i(\gamma_j)} \sim
P^{(\sum_{i,j} d_i(\gamma_j) +1)/2}. \ee Clearly, it would be nice
to study this conjecture further. Here we will see that it
provides a self-consistent picture of one-point functions in
various situations, one that moreover agrees with supergravity
calculations in cases where these are available.

\subsection{One-point functions}

We will now use the above results to find the large $N$ behavior
of correlation functions in various ensembles. Notice that most
results given here carry over to typical states in the ensembles
as well, provided (\ref{aux13}) still holds for the typical state.

\noindent\underline{$M=0$ BTZ} \\
\noindent We already have all the ingredients in place to do this
calculation. The ensemble consists of random cycles of total
length $N$. This is a microcanonical point of view, from a
canonical point of view it is perhaps better to think of it as a
system of 24 free bosons at a finite temperature proportional to
$N^{1/2}$. Either way, using (\ref{comb}) and (\ref{aux13}) we
find that \be \langle {\cal O}_k \rangle \sim
N^{1-g-\frac{1}{2}\sum_{i,j} d_i(\gamma_j)}. \ee The equation for
the genus (\ref{genus}) shows that $k$ has to be odd, otherwise
the one-point function is identically zero. It also shows that the
leading contribution comes from $g=0$ and we finally get \be
\label{finalans1} \langle {\cal O}_k \rangle \sim
N^{\frac{3-k}{4}}. \ee
In other words, to leading order all one-point functions of twist fields
vanish, but at subleading order in the $\alpha'\sim1/\sqrt{N}$ expansion
they are potentially turned on. It would be interesting to verify this
more explicitly using higher order curvature corrections to supergravity.

\noindent\underline{Small black ring, $q_3=1$} \\
\noindent For the small black ring, we separate the state into a
condensate $(\alpha^+_{-1})^J$ and a randomly distributed piece of
total length $P=N-J$. We denote by $\hat{d}$ the number of
elements of the condensate that live on the irreducible component
$\Sigma_0$. Also, by $\sum_{i,j}'$ we denote a sum over all cycles
except the condensate. Then the combinatorical factor works out to
be \be N^{\frac{1}{2} - g- \sum_{i,j}'d_i(\gamma_j) } \left(
\frac{J}{N} \right)^{\hat{d}} \frac{T}{R}
(N-J)^{(1+\sum_{i,j}'d_i(\gamma_j))/2}. \ee The leading
contribution appears when $g=0$, $\sum_{i,j}' d_i(\gamma_j)=1$ and
$\hat{d}=(k-1)/2$. In this case $\Sigma_0$ contains many states
from the condensate and only three cycles of length larger than
one. One can also check (this requires a bit of work) that for
these configurations $T/R=1$. Therefore, we finally obtain that
for odd $k$ \be \langle {\cal O}_k \rangle \sim N^{-1/2} \left(
\frac{J}{N} \right)^{\frac{k-1}{2}} (N-J) \label{aux21} . \ee When
we compare this to (\ref{finalsugra}) we need to take into account
that the $k$ used here is not the same as the $k$ used in
(\ref{finalsugra}). The supergravity fields used in the
calculations in sections~2,3 couple to chiral primary fields with
conformal weights $(\frac{k}{2},\frac{k}{2})$. One can check that
these are twisted fields that arise from the identity element of
$H^{\ast}(K3)$ in the $\mathbb Z_{k+1}$-twisted sector; therefore,
in order to compare (\ref{aux21}) to (\ref{finalsugra}) we should
replace $k$ by $k+1$ in (\ref{aux21}). After this substitution it
indeed agrees perfectly with (\ref{finalsugra}).

For $J\sim \sqrt{N}$ there is no longer a condensate, and we
expect the CFT result to scale in the same way as the $M=0$ BTZ.
This is indeed the case, which shows that our assumptions, in
particular (\ref{aux13}), are at least self-consistent.

\noindent\underline{Conical defect} \\
\noindent The conical defect has been conjectured
\cite{{Lunin:2001jy},{Lunin:2002iz}} to be dual to a pure state
$|A\rangle=(\alpha_{-p})^{N/p}$. The irreducible component will
therefore be of size $s=up$ for some $u$, and contain precisely
$u$ cycles of length $p$ from $|A\rangle$. We again expect the
leading contribution to come from genus zero surfaces, which
implies $k=2u-1$ (see (\ref{genus})). One may check however that
there are no such genus zero configurations. The cycle of length
$k$ would have exactly one element in common with at least one of
the cycles of length $p$ coming from $A$. The product of the two
will then involve a cycle of length larger than $p$, which is
inconsistent with the form of the state $|A\rangle$.

It therefore appears that the leading contributing is from genus
one, with $k=2u+1$, and one can explicitly find corresponding explicit group
elements that obey (\ref{relg}). With some more work we found that $T/R=u$
for this genus one case.

The precise combinatorial factor is then \be K=\left( \frac{(N-k)!
k}{N!} \right)^{1/2} u \frac{ p^u (N/p)!}{(N/p-u)!} . \ee
Amazingly, this implies that \be \langle {\cal O}_k \rangle \sim
\left(\frac{k-1}{2}\right) N^{-\frac{1}{2}} \ee whose
$N$-dependence is universal and independent of $k$. It would be
interesting to understand how this could arise from a supergravity
solution in the presence of  higher order $\alpha'$ corrections.

\noindent\underline{Small black ring, $q_3>1$. } \\
\noindent This case is somewhat similar to the small black ring
with $q_3=1$. The main difference is that now the condensate is
made up out of $(\alpha^+_{-q_3})^J$ instead of $(\alpha^+_{-1})^J$, and that the thermal
distribution has length $P=N-q_3 J$. The leading contribution is
from the same configuration as for the case $q_3=1$, namely
$\hat{d}=(k-1)/2$, $g=0$ and $\sum_{i,j}'d_i(\gamma_j)=1$. There
could in this case also have been a contribution with
$\sum_{i,j}'d_i(\gamma_j)=0$ which cannot happen for $q_3=1$, but
in view of our conical defect discussion this one would have to
have at least $g=1$ and is therefore subleading. Putting in all
the factors we get for odd $k$ that \be \langle {\cal O}_k \rangle
\sim N^{-1/2} \left( \frac{q_3J}{N} \right)^{\frac{k-1}{2}} (N-q_3
J) \ee which agrees (after shifting $k\rightarrow k+1$ as
explained above) with the supergravity result (\ref{finalsugra}), up
to a factor of $q_3^k$. Such a factor could well arise from the
CFT calculation once actual correlation functions are included, but this
is beyond the scope of the present calculation. It is already
quite nontrivial that the large $N,J$ dependence of the CFT and
supergravity results is identical.

\section{A toy model for higher condensation}
\label{toymodel}

The statistical mechanics of the ensemble corresponding to small
black rings, in absence of dipole charges, {\it i.e.} $q_3=1$, can
be studied by considering a partition function of the form ${\cal
Z}=Tr (e^{-\beta(H+\mu J)})$, where for the oscillators
$\alpha_{-n}^{\pm}$ $H=n$ and $J=\pm1$, while for $\alpha^i_{-n}$ with $i=3\ldots
24$ $H=n$ and $J=0$. It was shown in
\cite{Russo:1994ev,Iizuka:2005uv,Balasubramanian:2005qu} that in the
regime $J \sim N$ a Bose-Einstein condensate forms and the
ensemble consists of $J$ strings of length $1$ plus a thermal
distribution giving rise to the entropy $S \sim \sqrt{N-J}$.

In this section we study the partition function in presence of a
"dipole" charge $D$, {\it i.e.} ${\cal Z}=Tr (e^{-\beta(H+\mu
J+\nu D)})$. We show that if we assign to a given oscillator
$\alpha_{-n}^+$ a dipole charge $D=1/n$ then in a suitable regime the
expectation values of $J,D$ will be of order
$N$ and a Bose-Einstein condensation takes place but now of string
components of length $q_3 \geq 1$, furthermore we exactly
reproduce the small black ring entropy in the presence of dipoles
charges $S \sim \sqrt{N-q_3 J}$.

The partition function we want to study is given by
\begin{equation}
{\cal Z}=Tr_{\cal H} (e^{-\beta(H+\mu J+\nu D)})
\end{equation}
The Hilbert space ${\cal H}$ consists of a Fock space
built out of 24 free oscillators $\alpha^{\pm}_{-n}$ and $\alpha^i_{-n}$,
$i=3,...,24$, carrying the following charges \footnote{Note that
this assignment for the dipole charge D exactly coincides with
(4.18) of \cite{Alday:2003zb} which gives the action of the first
non-local charge of the infinite tower found in \cite{Bena:2003wd}
when acting on a BMN state.}
\begin{eqnarray}
[H,\alpha^{\pm}_{-n}]=n~\alpha^{\pm
}_{-n},\hspace{0.2in}[H,\alpha^i_{-n}]=n~\alpha^i_{-n}\\
~[J,\alpha^{\pm}_{-n} ] =\pm \alpha^{\pm
}_{-n},\hspace{0.2in}[J,\alpha^i_{-n}]=0,\hspace{0.2in}[D,\alpha^{+
}_{-n}]=\frac{1}{n}~\alpha^{+ }_{-n}
\end{eqnarray}
The charge of the other oscillators with respect to $D$ will not
be relevant for the discussion below, but will be relevant for
the subleading behavior of the entropy. This is discussed further in
the conclusions.

Let us focus on the
$\alpha^+$ oscillator, its contribution to the partition function
is
\begin{equation}
\log{\cal Z}=-\sum_{n=1}^{\infty} \log{(1-e^{\beta(-n
+\mu+\nu/n)})}=\sum_{n=1}^{\infty} {\cal C}_n
\end{equation}
and ${\cal C}_n$ can be written as
\begin{equation}
{\cal C}_n=\sum_{l=1}^\infty \frac{e^{-l n \beta}}{l} \left(
\sum_{j,k=0}^\infty \frac{(\mu \beta l)^j (l \beta \nu/n)^k}{j!
k!} \right) = \sum_{j,k=0}^\infty \frac{\mu^j \nu^k
\beta^{k+j}}{j! k! n^k} Li_{1-j-k}(e^{-\beta n}).
\end{equation}
After changing variables $k+j=s$ and summing over $0 \leq j \leq
s$ we get
\begin{equation}
{\cal C}_n=\sum_{s=0}^{\infty} \beta^s Li_{1-s}(e^{-\beta n})
\frac{(\nu +n \mu)^s}{n^s \Gamma(1+s)}
\end{equation}
Up to this point the above computation is exact, in order to
proceed we approximate in the limit $\beta \ll 1$ the polylogarithm
$Li_{1-s}$ for $s\geq 1$ by
\begin{equation}
Li_{1-s}(e^{-\beta n}) \approx \frac{(s-1)!}{\beta^s n^s}.
\end{equation}
Then
\begin{equation}
\tilde{{\cal C}}_n=\sum_{s=1}^{\infty} \beta^s Li_{1-s}(e^{-\beta
n}) \frac{(\nu +n \mu)^s}{n^s \Gamma(1+s)} \approx -
\log{(1-\frac{\mu}{n}-\frac{\nu}{n^2})}.
\end{equation}
The contribution from $s=0$ can be taken care of separately and the
sum over $n$ can easily be performed and gives the usual term
depending only on $\beta$. Taking into account all the oscillators
we get
\begin{equation} \label{zapp}
\log{{\cal Z}} \approx \frac{4 \pi^2}{\beta} -\sum_{n=1}^{\infty}
\log{(1-\frac{\mu}{n}-\frac{\nu}{n^2})}.
\end{equation}
The first term here is obtained by summing over all 24 oscillators,
but the second term is due only to $\alpha^+$. There are similar $\mu,\nu$-dependent
terms for the other oscillators as well, but
the reason for not including their contribution will become clear momentarily.
Computing the level $N$, the average angular momenta $J$ and the
average dipole charge $D$ from (\ref{zapp}) we get
\begin{eqnarray}
N=-\left( \frac{\partial \log{{\cal Z}}}{\partial \beta}
\right)_{\beta \mu,\beta \nu}=\frac{4
\pi^2}{\beta^2}+\mu J +\nu D \\
J= \left( \frac{\partial \log{{\cal Z}}}{\partial \beta \mu}
\right)_{\beta,\beta \nu} =\sum_{n=1}^{\infty} \frac{n}{n^2 \beta-n \beta \mu-\beta \nu}\\
D= \left( \frac{\partial \log{{\cal Z}}}{\partial \beta \nu}
\right)_{\beta,\beta \mu}=\sum_{n=1}^{\infty} \frac{1}{n^2 \beta-n
\beta \mu-\beta \nu}.
\end{eqnarray}
The expression for $J$ appears to diverge, but that is due to the
approximation that we made. If we include the contribution from
$\alpha^-$, which is similar to that of $\alpha^+$ in (\ref{zapp})
except that $\mu$ is replaced by $-\mu$, the expression for $J$
will be convergent. This $\alpha^-$ contribution will not be
relevant for most of what follows though. The expressions for
$J,D$ are at first sight of order $\sqrt{N}\sim\beta^{-1}$. To see
this we need to include the contribution from $\alpha^-$ in $J$.
In order for $J,D$ to be of order $N$, one term in the sum must be
very large; if this happens for the term with $n=q_3$ then in
order to have $J, D \sim N$ we need that
\begin{equation}
\label{pole}
q_3^2 - q_3 \mu - \nu \sim \beta \ll 1
\end{equation}
Notice that this will imply condensation of modes with $n=q_3$,
indeed
\begin{equation}
<0|\alpha_{-n}^{+} \alpha_n^+|0> = \frac{e^{\beta(-n
+\mu+\nu/n)}}{1-e^{\beta(-n +\mu+\nu/n)}} ,
\end{equation}
which has a pole at $n=q_3$ for $q_3^2 - q_3 \mu - \nu=0$.
Obviously, the combination $n^2 -n \mu -\nu$ has to be greater
than 0 for all $n$ otherwise the thermodynamic system is
ill-defined. If we also require that this quantity has a minimum
obeying (\ref{pole}) at $n=q_3$, we find
\begin{equation}
\mu \approx 2 q_3 ,\hspace{0.3in}\nu \approx -q_3^2.
\end{equation}
With these values of $\mu,\nu$ the term with $n=q_3$ will dominate
the sum that appears in the partition function in (\ref{zapp}). If
we keep only that term together with the other contribution
$4\pi^2/\beta$ we can compute the entropy and we find
\begin{equation}
S=\beta (N-\mu J -\nu D)+ \log{\cal Z} \approx \frac{8
\pi^2}{\beta} = 4 \pi \sqrt{N -\mu J -\nu D}=4 \pi \sqrt{N-q_3 J}.
\end{equation}
That agrees exactly with the (small) black ring entropy for a
general dipole charge $q_3$!

\section{Conclusions}

Black rings have many features that make them interesting objects
to study,  in particular they provide examples of gravitational
solutions that on the one hand carry a non trivial dipole charge
and on the other hand admit a description in terms of a dual CFT.
In this paper we have studied several aspects of this system, in
particular we tried to understand the nature of the dipole charge
from the dual CFT perspective.

Using standard AdS/CFT one can extract the one point functions of
CFT operators in the ensemble dual to the black ring. In this
paper we have focussed on scalar operators, to complete the
analysis one should consider vector and tensor operators as well,
which is in principle straightforward but rather tedious. The one
point functions are complicated expressions of the seven
parameters of the solution. Once we restrict to the case of
$1/2$-BPS or small black rings they simplify considerably (as many
of the parameters are set to zero). The small black ring has
vanishing macroscopic entropy, however we have shown that choosing
appropriately the value of some parameters in the full
seven-parameter solution both the microscopic entropy and one
point functions of the small black ring can be reproduced. This
would suggest these solutions have similarities with the small
black ring once stringy corrections are included. It would be
interesting to explore this further. As we have worked in the
limit of large charges, such a conjecture could be further
tested/refined by studying subleading corrections to the entropy
and one point functions. Another interesting question would be to
understand whether these small extra charges, which are introduced
by hand but strictly speaking absent in the small black ring, can
be understood as arising from some polarization effect.

We have also computed the one point functions directly in the
orbifold CFT in the type of ensembles that are believed to be dual
to the $1/2-BPS$ black ring. The computation is very complicated
to perform and we need to make a number of assumptions (in
particular eq. (\ref{aux13})) in order to proceed; it would be
nice to test further the validity of our assumptions. We have
found that the leading $N$ contribution to the one point function
for the small black ring agrees with those computed from
supergravity. As a by-product we have found that one point
functions in conical defects are subleading in $1/N$ but there is
no obvious reason why they should vanish. It would be interesting
to see whether this can be reproduced from supergravity once
higher $\alpha'$ corrections are included.

In section~5 we have presented a simple toy model that correctly
seems to reproduce the physics of the small black ring. To a
string of length $n$ we associate a "dipole charge" $D=1/n$, which
indicates that the dipole charge corresponds to some non-local
operator in the CFT; this is perhaps exactly what one expects for
a dipole charge. Our proposal can easily be generalized (by
including further negative powers of $n$) to more general
solutions, for instance concentric black rings, and it would be
interesting to study this in more detail. This same kind of
expressions appear in many places in integrable systems. For
instance, when considering the tower of non-local charges for
strings on $AdS_5 \times S^5$ acting on BMN states the first non
local charge has this same expression. This suggests that it may
be worthwhile to study a thermodynamic system of strings in AdS
which includes a potential for each of the non-local charges,
perhaps using the integrable model approach to strings in AdS.

Finally it remains an open problem to study the phase diagram of
the thermodynamic toy model, and the corresponding supergravity
solutions. For example, the toy model includes the description of
conical defects. There is no reason, in the toy model, why $q_3$
should be an integer, and allowing arbitrary real $q_3$ could
perhaps lead to families of supergravity solutions that
interpolate between different conical defects. We should also
point out that in general it is not clear exactly what type of
ensemble one should use in general to describe gravitational
solutions. Different ensembles usually yield the same leading
answer for the entropy, but different subleading pieces. In
\cite{dabholkarrecent} it was argued that small black rings should
not be described as an ensemble with fixed angular momentum and
energy, but rather as a system with a fixed condensate and energy.
Both points of view yield different subleading terms in the
expansion of the entropy in large charges. It would be interesting
to understand the connection between this suggestion and our toy
model.

\section*{Acknowledgments}

\noindent We would like to thank Iosif Bena, Justin David, Per
Kraus, Finn Larsen, Donald Marolf and Pedro Silva for useful
discussions, and Vijay Balasubramanian and Roberto Emparan for
useful discussions and insightful comments on a draft of this
paper. This work was supported in part by the stichting FOM.

\appendix

\section{$SL(2,\mathbb R) \times SL(2,\mathbb R)$ generators}

\subsection{$M=0$ BTZ generators}
\label{m0vir}

The isometry group of $AdS_3$, $SL(2,\mathbb R)_L \times
SL(2,\mathbb R)_R$ is generated by the Virasoro generators $L_0,
L_{\pm 1}$ and $\bar{L}_0,\bar{L}_{\pm 1}$. In cylindrical
coordinates the $AdS_3$ metric reads
\begin{equation}
\label{cyl}
 \frac{ds^2}{Q}=-\cosh^2 \rho d \tau^2+\sinh^2 \rho d
\phi^2 +d \rho^2
\end{equation} and the Virasoro generators are given by
\cite{Maldacena:1998bw}\cite{deBoer:1998ip}
\begin{eqnarray}
L_0=i \partial_u,\\
L_{-1}=i e^{-i u} \left( \frac{\cosh 2 \rho}{\sinh 2
\rho}\partial_u-\frac{1}{\sinh 2 \rho} \partial_v+\frac{i}{2}\partial_\rho \right),\\
L_{1}=i e^{i u} \left( \frac{\cosh 2 \rho}{\sinh 2
\rho}\partial_u-\frac{1}{\sinh 2
\rho}\partial_v-\frac{i}{2}\partial_\rho \right)
\end{eqnarray}
with $u=\tau +\phi$, $v=\tau-\phi$. They satisfy the following
commutation relations
\begin{equation}
 [L_0,L_{\pm1}]=\mp L_{\pm1},\hspace{0.3in}[L_{1},L_{-1}]=2L_0.
\end{equation}
The right moving generators $\bar{L}_0$, $\bar{L}_{\pm 1}$ are
given by similar expressions with $u \leftrightarrow v$.

The change of coordinates from (\ref{cyl}) to the $M=0$ BTZ metric
is given by
\begin{eqnarray}
r&=&Q ( \cosh \rho \cos \tau+\sinh \rho \cos \phi),\\
t&=&\frac{Q}{r} \cosh \rho \sin \tau,\\
z&=&\frac{Q}{r}\sinh \rho \sin \phi.
\end{eqnarray}
From this one obtains that the local $SL(2,\mathbb R)_L$
generators of the $M=0$ BTZ metric are given by the vector fields
\begin{eqnarray} \label{lii}
L_0 & = & -\frac{1}{2} i r(t+z)\partial_r + \frac{i
(-Q^2+r^2(1+(t+z)^2))}{4
r^2}\partial_z+\frac{i(Q^2+r^2(1+(t+z)^2))}{4r^2} \partial_t \nonumber\\
L_1 & = & \frac{1}{2}i
r(-i+t+z)\partial_r+\frac{i(Q^2-r^2(-i+t+z)^2)}{4r^2}\partial_z-\frac{i(Q^2+r^2(-i+t+z)^2)}{4r^2}\partial_t \nonumber \\
L_{-1}& = & \frac{1}{2}i
r(i+t+z)\partial_r+\frac{i(Q^2-r^2(i+t+z)^2)}{4r^2}\partial_z-\frac{i(Q^2+r^2(i+t+z)^2)}{4r^2}\partial_t.
\end{eqnarray}
The $SL(2,R)_R$ generators are obtained simply by taking $z
\rightarrow -z$. The corresponding quadratic Casimirs are given by
\begin{eqnarray}
L^2=\frac{1}{2}(L_{-1}L_{+1}+L_{+1}L_{-1})-L_0^2\\
\bar{L}^2=\frac{1}{2}(\bar{L}_{-1}\bar{L}_{+1}+\bar{L}_{+1}\bar{L}_{-1})-\bar{L}_0^2.
\end{eqnarray}
Of course, the $M=0$ BTZ metric does not have a global
$SL(2,\mathbb R)$ isometry group because it is a quotient of
global AdS${}_3$. This is reflected in (\ref{lii}) by the fact
that these generators are not globally well-defined, due to the
periodicity of the coordinate $z$. Still, the quadratic Casimirs
are well-defined, and they provide the kinetic terms for the
various fields that propagate in the $M=0$ BTZ background.

\subsection{Asymptotic Virasoro generators}

\label{mm1vir}

As discussed in section~2, when we compute the one-point functions
we should really view the large black ring as a ``large''
perturbation of global AdS${}_3$, not as a ``small'' perturbation
of the $M=0$ BTZ black hole. Thus, the right Virasoro generators
we should use are those of AdS${}_3$ written in coordinates in
which the asymptotic behavior is identical to that of the black
ring solution, which in turn is identical to asymptotic behavior
of the $M=0$ BTZ solution. The right coordinates are those in
which we view global AdS${}_3$ as the $M=-1$ BTZ solution. They
can be obtained from (\ref{cyl}) through the change of coordinates
\begin{eqnarray}
\sinh \rho=r/Q
\end{eqnarray}
and the Virasoro generators then become (in $r,t=\tau,z=\phi$
coordinates)
\begin{eqnarray}
L_0& =& \frac{i}{2}(\partial_t+\partial_z)\\
L_{+}& =& \frac{i}{2}e^{i(t+z)}(-i\sqrt{Q^2+r^2}\partial_r+
\sqrt{1+Q^2/r^2}
\partial_z+\frac{1}{\sqrt{1+Q^2/r^2}}\partial_t)\\
L_{-}& =& \frac{i}{2}e^{-i(t+z)}(i\sqrt{Q^2+r^2}\partial_r+
\sqrt{1+Q^2/r^2} \partial_z+\frac{1}{\sqrt{1+Q^2/r^2}}\partial_t)
\end{eqnarray}
with analogous expressions for the right-moving generators, with
$z \rightarrow -z$. Notice that these generators are globally well
defined. On the other hand, they are exact isometries of the
$M=-1$ BTZ metric (i.e. global AdS${}_3$), but only asymptotic (at
large $r$) isometries of the $M=0$ BTZ.

\section{$SO(4)$ harmonics}

In terms of embedding coordinates $X^0,...,X^3$ the $S^3$ metric
is written as
\begin{equation}
ds^2=(dX^0)^2+...+(dX^3)^2.
\end{equation}
The isometry group of $S^3$ is $SO(4)$, whose generators in terms
of the embedding coordinates $X^0,...,X^3$ are given by
\begin{equation}
L_{ab}=X^a \partial_b -X^b \partial_a
\end{equation}
subject to the constraint $(X^0)^2+...+(X^3)^2=1$. The appropriate
change of coordinates to the metric given in $\ref{metricsol}$ is
\begin{eqnarray}
X^0=\cos\theta \cos\phi,\hspace{0.3in}X^1=\cos\theta \sin\phi,\\
X^2=\sin\theta \cos\psi,\hspace{0.3in}X^3=\sin\theta \sin\psi.
\end{eqnarray}
The group $SO(4)$ is isomorphic to two copies of $SU(2)$. One of
these is generated by
\begin{eqnarray}
G_1&=&\frac{L_{23}-L_{01}}{2}=\frac{1}{2}( \partial_\psi -\partial_\phi)\\
G_2&=&\frac{L_{12}-L_{03}}{2}= \frac{1}{2} \left( -\cot \theta
\cos{(\phi-\psi)} \partial_\psi
-\tan \theta \cos(\phi-\psi) \partial_\phi + \sin(\phi-\psi) \partial_\theta \right)\\
G_3&=&\frac{L_{02}-L_{13}}{2}=\frac{1}{2} \left( \cot \theta
\sin{(\phi-\psi)} \partial_\psi +\tan \theta \sin(\phi-\psi)
\partial_\phi + \cos(\phi-\psi) \partial_\theta \right)
\end{eqnarray}
and the other one with generators $\bar{G}_i$ is generated by the
same vector fields but with $\phi$ replaced by $ -\phi$. They
satisfy the following commutation relations
\begin{equation}
 [G_1,G_2]=-G_3,\hspace{0.3in}
 [G_2,G_3]=-G_1,\hspace{0.3in}[G_3,G_1]=-G_2
\end{equation}
together with $[G_i,\bar{G}_j]=0$. Notice that the perturbation is
trivially invariant under the action of $G_1$ and $\bar{G}_1$ that
we will take as the Cartan generators. The quadratic Casimirs are
given by
\begin{eqnarray}
G^2=-4 \left( G_1^2+G_2^2+G_3^2\right)\\
\bar{G}^2=-4 \left(\bar{G}_1^2+\bar{G}_2^2+\bar{G}_3^2\right).
\end{eqnarray}
The scalar spherical harmonics are given by
\begin{equation}
Y^{k}_s(\theta)=\sqrt{1+k}~F(1+\frac{k}{2},-\frac{k}{2};1,\sin^2\theta)
\end{equation}
for $k=0,2,4,6,...$. The action of the quadratic Casimirs on the
spherical harmonics is as follows
\begin{equation}
G^2 Y^{k}_s(\theta)=\bar{G}^2
Y^{k}_s(\theta)=k(k+2)Y^{k}_s(\theta),
\end{equation}
where the normalization constant has been chosen so that
\begin{equation}
\int_{S^3} Y^{k}_s Y^{k'}_s=\int_0^{\pi/2} 2\sin\theta \cos\theta
Y^{k}(\theta)_s Y^{k'}(\theta)_s d \theta=\delta^{k k'}.
\end{equation}
Notice that with this normalization
\begin{eqnarray}
\int_{S^3}\nabla^a Y^{k}_s \nabla_a Y^{k'}_s& =& k(k+2)\delta^{k k'}\\
\int_{S^3} \nabla^{(a} \nabla^{b)} Y^{k}_s \nabla_a \nabla_b
Y^{k'}_s& =& \frac{2}{3}k(k+2)(k(k+2)-3)\delta^{k k'}
\end{eqnarray}
where we use $(a,b)$ to denote the symmetric traceless combination
and indices are raised and lowered with the $S^3$ metric
$G^{0}_{ab}=Diag(\sin^2\theta,\cos^2\theta,1)$.

The vector spherical harmonics are given by
\begin{equation}
Y_{v \pm}^k(\theta)=\sqrt{\frac{k}{2}}~\left(~
-\frac{k}{2}\sin^2\theta
F(1-\frac{k}{2},1+\frac{k}{2};2,\sin^2\theta)~~,~~\pm
F(-\frac{k}{2},\frac{k}{2};1,\sin^2\theta)~~,~~0~\right)
\end{equation}
for $k=2,4,6,...$. The quadratic Casimirs act as follows
\begin{equation}
G^2 Y_{v \pm}^k = (k-1 \pm 1)(k+1 \pm 1) Y_{v
\pm}^k,\hspace{0.3in} \bar{G}^2 Y_{v \pm}^k=(k-1 \mp 1)(k+1 \mp 1)
Y_{v \pm}^k
\end{equation}
The normalization factor has been chosen so that
\begin{equation}
\int_{S^3} (G^0)^{a b} (Y_{v \pm}^k)_a (Y_{v
\pm}^{k'})_b=\delta^{k k'}
\end{equation}

Finally, the tensor spherical harmonics are given by
\begin{equation}
Y_{t \pm}^k=\frac{1}{4}\sqrt{k(k-1)(k-2)}\left(
\begin{array}{ccc}
  f_1(\theta) & \pm g(\theta) & 0  \\
  \pm g(\theta) & f_2(\theta) & 0  \\
  0 & 0 & f_3(\theta)
\end{array} \right)
\end{equation}
with
\begin{eqnarray}
f_1(\theta)&=&\frac{1}{4(k-1)}\left( -4\sin^2\theta
F(1-\frac{k}{2},\frac{k}{2};2,\sin^2\theta)+k(k-2)\sin^4\theta
F(2-\frac{k}{2},1+\frac{k}{2};3,\sin^2\theta)\right) \nonumber\\
f_2(\theta)&=& \frac{F(1-\frac{k}{2},\frac{k}{2};2,\sin^2\theta)}{k-1}-\cot^2{\theta}f_1(\theta) \nonumber\\
f_3(\theta)&=&
\frac{F(1-\frac{k}{2},\frac{k}{2};2,\sin^2\theta)}{(1-k)\cos^2\theta}
\nonumber\\
g(\theta)&=&\sin^2\theta
F(1-\frac{k}{2},\frac{k}{2};2,\sin^2\theta).
\end{eqnarray}
The action of the quadratic casimirs is given by
\begin{equation}
G^2 Y_{t \pm}^k = (k-2 \pm 2)(k \pm 2) Y_{t \pm}^k,\hspace{0.3in}
\bar{G}^2 Y_{t \pm}^k=(k-2 \mp 2)(k \mp 2) Y_{t \pm}^k.
\end{equation}
The normalization factor has been chosen so that
\begin{equation}
\int_{S^3} (G^0)^{a c} (G^0)^{b d} (Y_{t \pm}^k)_{a b} (Y_{t
\pm}^{k'})_{c d}=\delta^{k k'}.
\end{equation}



\begin{thebibliography}{99}

\bibitem{Maldacena:2001kr}
  J.~M.~Maldacena,
  JHEP {\bf 0304}, 021 (2003)
  [arXiv:hep-th/0106112].

\bibitem{Lunin:2001jy}
  O.~Lunin and S.~D.~Mathur,
  ``AdS/CFT duality and the black hole information paradox,''
  Nucl.\ Phys.\ B {\bf 623} (2002) 342
  [arXiv:hep-th/0109154].

\bibitem{mathur2}
  O.~Lunin, S.~D.~Mathur and A.~Saxena,
  ``What is the gravity dual of a chiral primary?,''
  Nucl.\ Phys.\ B {\bf 655}, 185 (2003)
  [arXiv:hep-th/0211292].


\bibitem{Mathur:2005zp}
  S.~D.~Mathur,
  ``The fuzzball proposal for black holes: An elementary review,''
  Fortsch.\ Phys.\  {\bf 53} (2005) 793
  [arXiv:hep-th/0502050].

\bibitem{Lin:2004nb}
  H.~Lin, O.~Lunin and J.~Maldacena,
  ``Bubbling AdS space and 1/2 BPS geometries,''
  JHEP {\bf 0410} (2004) 025
  [arXiv:hep-th/0409174].

\bibitem{ma1}
  A.~Buchel,
  ``Coarse-graining 1/2 BPS geometries of type IIB supergravity,''
  arXiv:hep-th/0409271.

\bibitem{ma2}
  N.~V.~Suryanarayana,
  ``Half-BPS giants, free fermions and microstates of superstars,''
  arXiv:hep-th/0411145.

\bibitem{ma3}
  P.~G.~Shepard,
  ``Black hole statistics from holography,''
  JHEP {\bf 0510}, 072 (2005)
  [arXiv:hep-th/0507260].

\bibitem{Balasubramanian:2005mg}
  V.~Balasubramanian, J.~de Boer, V.~Jejjala and J.~Simon,
  ``The library of Babel: On the origin of gravitational thermodynamics,''
  arXiv:hep-th/0508023.

\bibitem{ma4}
  P.~J.~Silva,
  ``Rational foundation of GR in terms of statistical mechanic in the AdS/CFT
  framework,''
  arXiv:hep-th/0508081.


\bibitem{Myers:2001aq}
  R.~C.~Myers and O.~Tafjord,
  ``Superstars and giant gravitons,''
  JHEP {\bf 0111} (2001) 009
  [arXiv:hep-th/0109127].


\bibitem{Emparan:2004wy}
  R.~Emparan,
  ``Rotating circular strings, and infinite non-uniqueness of black rings,''
  JHEP {\bf 0403} (2004) 064
  [arXiv:hep-th/0402149].

\bibitem{Copsey:2005se}
  K.~Copsey and G.~T.~Horowitz,
  ``The role of dipole charges in black hole thermodynamics,''
  arXiv:hep-th/0505278.


\bibitem{Corley:2001zk}
  S.~Corley, A.~Jevicki and S.~Ramgoolam,
  ``Exact correlators of giant gravitons from dual N = 4 SYM theory,''
  Adv.\ Theor.\ Math.\ Phys.\  {\bf 5} (2002) 809
  [arXiv:hep-th/0111222].

\bibitem{davber}
D.~Berenstein,
  ``A toy model for the AdS/CFT correspondence,''
  JHEP {\bf 0407}, 018 (2004)
  [arXiv:hep-th/0403110].

\bibitem{Elvang:2004ds}
  H.~Elvang, R.~Emparan, D.~Mateos and H.~S.~Reall,
  ``Supersymmetric black rings and three-charge supertubes,''
  Phys.\ Rev.\ D {\bf 71} (2005) 024033
  [arXiv:hep-th/0408120].


\bibitem{Bena:2004tk}
  I.~Bena and P.~Kraus,
  ``Microscopic description of black rings in AdS/CFT,''
  JHEP {\bf 0412} (2004) 070
  [arXiv:hep-th/0408186].

\bibitem{Iizuka:2005uv}
  N.~Iizuka and M.~Shigemori,
  ``A note on D1-D5-J system and 5D small black ring,''
  JHEP {\bf 0508} (2005) 100
  [arXiv:hep-th/0506215].

\bibitem{Balasubramanian:2005qu}
  V.~Balasubramanian, P.~Kraus and M.~Shigemori,
  ``Massless black holes and black rings as effective geometries of the D1-D5
  system,''
  arXiv:hep-th/0508110.


\bibitem{don}B.~C.~Palmer and D.~Marolf,
  ``Counting supertubes,''
  JHEP {\bf 0406}, 028 (2004)
  [arXiv:hep-th/0403025].

\bibitem{dabholkarrecent}
  A.~Dabholkar, N.~Iizuka, A.~Iqubal and M.~Shigemori,
  ``Precision microstate counting of small black rings,''
  arXiv:hep-th/0511120.



\bibitem{Balasubramanian:2000rt}
  V.~Balasubramanian, J.~de Boer, E.~Keski-Vakkuri and S.~F.~Ross,
  ``Supersymmetric conical defects: Towards a string theoretic description  of
  black hole formation,''
  Phys.\ Rev.\ D {\bf 64} (2001) 064011
  [arXiv:hep-th/0011217].

\bibitem{Maldacena:2000dr}
  J.~M.~Maldacena and L.~Maoz,
  ``De-singularization by rotation,''
  JHEP {\bf 0212} (2002) 055
  [arXiv:hep-th/0012025].



\bibitem{Lunin:2002iz}
  O.~Lunin, J.~Maldacena and L.~Maoz,
  ``Gravity solutions for the D1-D5 system with angular momentum,''
  arXiv:hep-th/0212210.

\bibitem{Dabholkar:2004yr}
  A.~Dabholkar,
  ``Exact counting of black hole microstates,''
  Phys.\ Rev.\ Lett.\  {\bf 94} (2005) 241301
  [arXiv:hep-th/0409148].


\bibitem{Gauntlett:2004wh}
  J.~P.~Gauntlett and J.~B.~Gutowski,
  ``Concentric black rings,''
  Phys.\ Rev.\ D {\bf 71} (2005) 025013
  [arXiv:hep-th/0408010].

\bibitem{Gauntlett:2004qy}
  J.~P.~Gauntlett and J.~B.~Gutowski,
  ``General concentric black rings,''
  Phys.\ Rev.\ D {\bf 71} (2005) 045002
  [arXiv:hep-th/0408122].


\bibitem{Alday:2003zb}
  L.~F.~Alday,
  ``Non-local charges on AdS(5) x S**5 and pp-waves,''
  JHEP {\bf 0312} (2003) 033
  [arXiv:hep-th/0310146].

\bibitem{Bena:2003wd}
  I.~Bena, J.~Polchinski and R.~Roiban,
  ``Hidden symmetries of the AdS(5) x S**5 superstring,''
  Phys.\ Rev.\ D {\bf 69} (2004) 046002
  [arXiv:hep-th/0305116].


\bibitem{Elvang:2004rt}
  H.~Elvang, R.~Emparan, D.~Mateos and H.~S.~Reall,
  ``A supersymmetric black ring,''
  Phys.\ Rev.\ Lett.\  {\bf 93} (2004) 211302
  [arXiv:hep-th/0407065].

\bibitem{Breckenridge:1996is}
  J.~C.~Breckenridge, R.~C.~Myers, A.~W.~Peet and C.~Vafa,
  ``D-branes and spinning black holes,''
  Phys.\ Lett.\ B {\bf 391} (1997) 93
  [arXiv:hep-th/9602065].


\bibitem{Emparan:2005jk}
  R.~Emparan and D.~Mateos,
  ``Oscillator level for black holes and black rings,''
  Class.\ Quant.\ Grav.\  {\bf 22} (2005) 3575
  [arXiv:hep-th/0506110].


\bibitem{Deger:1998nm}
  S.~Deger, A.~Kaya, E.~Sezgin and P.~Sundell,
  ``Spectrum of D = 6, N = 4b supergravity on AdS(3) x S(3),''
  Nucl.\ Phys.\ B {\bf 536} (1998) 110
  [arXiv:hep-th/9804166].

\bibitem{Arutyunov:2000by}
  G.~Arutyunov, A.~Pankiewicz and S.~Theisen,
  ``Cubic couplings in D = 6 N = 4b supergravity on AdS(3) x S(3),''
  Phys.\ Rev.\ D {\bf 63} (2001) 044024
  [arXiv:hep-th/0007061].


\bibitem{Mihailescu:1999cj}
  M.~Mihailescu,
  ``Correlation functions for chiral primaries in D = 6 supergravity on  AdS(3)
  x S(3),''
  JHEP {\bf 0002} (2000) 007
  [arXiv:hep-th/9910111].


\bibitem{Freedman:1998tz}
  D.~Z.~Freedman, S.~D.~Mathur, A.~Matusis and L.~Rastelli,
  ``Correlation functions in the CFT($d$)/AdS($d+1$) correspondence,''
  Nucl.\ Phys.\ B {\bf 546} (1999) 96
  [arXiv:hep-th/9804058].

\bibitem{Klebanov:1999tb}
  I.~R.~Klebanov and E.~Witten,
  ``AdS/CFT correspondence and symmetry breaking,''
  Nucl.\ Phys.\ B {\bf 556} (1999) 89
  [arXiv:hep-th/9905104].

\bibitem{Russo:1994ev}
  J.~G.~Russo and L.~Susskind,
  ``Asymptotic level density in heterotic string theory and rotating black
  holes,''
  Nucl.\ Phys.\ B {\bf 437}, 611 (1995)
  [arXiv:hep-th/9405117].


\bibitem{Maldacena:1998bw}
  J.~M.~Maldacena and A.~Strominger,
  ``AdS(3) black holes and a stringy exclusion principle,''
  JHEP {\bf 9812} (1998) 005
  [arXiv:hep-th/9804085].

\bibitem{deBoer:1998ip}
  J.~de Boer,
  ``Six-dimensional supergravity on S**3 x AdS(3) and 2d conformal field
  theory,''
  Nucl.\ Phys.\ B {\bf 548} (1999) 139
  [arXiv:hep-th/9806104].

\bibitem{Strominger:1996sh}
  A.~Strominger and C.~Vafa,
  ``Microscopic Origin of the Bekenstein-Hawking Entropy,''
  Phys.\ Lett.\ B {\bf 379}, 99 (1996)
  [arXiv:hep-th/9601029].

\bibitem{Seiberg:1999xz}
  N.~Seiberg and E.~Witten,
  ``The D1/D5 system and singular CFT,''
  JHEP {\bf 9904} (1999) 017
  [arXiv:hep-th/9903224].

\bibitem{deBoer:1998us}
  J.~de Boer,
  ``Large N Elliptic Genus and AdS/CFT Correspondence,''
  JHEP {\bf 9905} (1999) 017
  [arXiv:hep-th/9812240].

\bibitem{Dijkgraaf:1998gf}
  R.~Dijkgraaf,
  ``Instanton strings and hyperKaehler geometry,''
  Nucl.\ Phys.\ B {\bf 543} (1999) 545
  [arXiv:hep-th/9810210].

\bibitem{Larsen:1999uk}
  F.~Larsen and E.~J.~Martinec,
  ``U(1) charges and moduli in the D1-D5 system,''
  JHEP {\bf 9906} (1999) 019
  [arXiv:hep-th/9905064].

\bibitem{David:1999ec}
  J.~R.~David, G.~Mandal and S.~R.~Wadia,
  ``D1/D5 moduli in SCFT and gauge theory, and Hawking radiation,''
  Nucl.\ Phys.\ B {\bf 564} (2000) 103
  [arXiv:hep-th/9907075];
 ``Microscopic formulation of black holes in string theory,''
  Phys.\ Rept.\  {\bf 369}, 549 (2002)
  [arXiv:hep-th/0203048].

\bibitem{Harvey:1995fq}
  J.~A.~Harvey and G.~W.~Moore,
  ``Algebras, BPS States, and Strings,''
  Nucl.\ Phys.\ B {\bf 463} (1996) 315
  [arXiv:hep-th/9510182].

\bibitem{Harvey:1996gc}
  J.~A.~Harvey and G.~W.~Moore,
  ``On the algebras of BPS states,''
  Commun.\ Math.\ Phys.\  {\bf 197} (1998) 489
  [arXiv:hep-th/9609017].

\bibitem{lehnsorger}
M.~Lehn, C.~Sorger, ``The cup product of the Hilbert scheme for K3
surface'' [arXiv:math.AG/0012166].

\bibitem{ruan}
Y.~Ruan, ``Stringy orbifolds'' [arXiv:math.AG/0201123].

\bibitem{wang}
W.~Wang, ``Universal rings arising in geometry and group theory''
[arXiv:math.QA/0211093].



\bibitem{Jevicki:1998bm}
  A.~Jevicki, M.~Mihailescu and S.~Ramgoolam,
  ``Gravity from CFT on S**N(X): Symmetries and interactions,''
  Nucl.\ Phys.\ B {\bf 577} (2000) 47
  [arXiv:hep-th/9907144].














\bibitem{Lunin:2001pw}
  O.~Lunin and S.~D.~Mathur,
  ``Three-point functions for M(N)/S(N) orbifolds with N = 4 supersymmetry,''
  Commun.\ Math.\ Phys.\  {\bf 227} (2002) 385
  [arXiv:hep-th/0103169].

\bibitem{Larsen:1998xm}
  F.~Larsen,
  ``The perturbation spectrum of black holes in N = 8 supergravity,''
  Nucl.\ Phys.\ B {\bf 536}, 258 (1998)
  [arXiv:hep-th/9805208].



\end{thebibliography}
\end{document}